\documentstyle[aps,url,prb]{revtex}
\begin{document}
\newcommand{\be}{\begin{equation}}
\newcommand{\ee}{\end{equation}}
\newcommand{\bea}{\begin{eqnarray}}
\newcommand{\eea}{\end{eqnarray}}
\newcommand{\Eq}[1]{ Eq.\,\,(\ref{#1})}
\newcommand{\Eqs}[1]{ Eqs.\ (\ref{#1})}
\newcommand{\Ref}[1]{Ref.~\onlinecite{#1}}
\newcommand{\ket}[1]{|#1 \rangle}
\draft
\title{Anharmonic force field and vibrational frequencies of
tetrafluoromethane (CF$_4$) and tetrafluorosilane (SiF$_4$)}
\author{Xiao-Gang Wang\thanks{Corresponding author for the vibrational 
calculation. Email: {\tt xgwang@emu.thchem.ox.ac.uk}}}
\address{Physical and Theoretical Chemistry Laboratory,
University of Oxford,
South Parks Road, Oxford OX1 3QZ, United Kingdom.}
\author{Edwin L. Sibert III}
\address{Department of Chemistry and Theoretical Chemistry Institute,
University of Wisconsin --- Madison, Madison, WI  53706, USA}
\author{Jan M. L. Martin\thanks{Corresponding author for the force field
calculation. Email: {\tt comartin@wicc.weizmann.ac.il}}}
\address{Department of Organic Chemistry,
Weizmann Institute of Science,
76100 Re\d{h}ovot, Israel}
\date{Submitted to {\it J. Chem. Phys.} September 17, 1999}

\maketitle

\begin{abstract}
Accurate quartic anharmonic force fields for CF$_4$ and SiF$_4$ have been calculated
using the CCSD(T) method and basis sets of $spdf$ quality.
Based on the {\it ab initio} force field with a minor empirical adjustment,
the vibrational energy levels of these two molecules and their isotopomers
are calculated by means of
high order Canonical Van Vleck Perturbation Theory(CVPT)
based on curvilinear coordinates.
The calculated energies agree very well with the experimental data.
The full quadratic force field of CF$_4$ is further refined to the
experimental data.
The symmetrization of the Cartesian basis for any combination bands
of $T_d$ group molecules
is discussed using the circular promotion
operator for the doubly degenerate modes,
together with tabulated vector coupling coefficients.
The extraction of the spectroscopic constants from our second order
transformed Hamiltonian in curvilinear coordinates
is discussed,
and compared to a similar procedure in rectilinear coordinates.
\end{abstract}

\section{Introduction}


Carbon tetrafluoride (also know as Freon-14) and tetrafluorosilane
are molecules of wide ranging industrial, environmental, and
economical interest.  While the use of carbon tetrafluoride is
being phased out together with that of other CFCs, it has a number
of other important industrial and technological
applications.\cite{KirkOthmer} These include use as an isolator
and extinguisher gas in high-voltage applications, and as an
etching gas in the semiconductor industry.\cite{KirkOthmer} The
formation of CF$_4$\ldots{}O$_2$ adducts was considered as a model
system for the use of liquid perfluorochemicals in artificial
blood.\cite{blood} Tetrafluorosilane, SiF$_4$, is a precursor
(e.g. by glow discharge in SiF$_4$-H$_2$ mixtures) of amorphous
Si-F-H semiconductors,\cite{KirkOthmer} as well as for plasma
deposition of low-dielectric Si-O-F thin solid films.\cite{Kim96}
It is the main by-product of beam etching of both semiconductors
with fluorine, and silicon dioxide with fluorocarbons.\cite{JAP}
In addition to being used to monitor the above processes [see e.g.
\Ref{philips}], SiF$_4$ has been proposed\cite{Francis} as
a remote volcano monitoring probe, since its presence can be
measured using open-path FT-IR spectroscopy of $\nu_3$(SiF$_4$).
Finally, the isotopic separation of $^{30}$Si (in $^{30}$SiF$_4$)
by infrared multi-photon dissociation (IRMPD) of natural-abundance
Si$_2$F$_6$ using CO$_2$ lasers has been reported on a preparative
scale.\cite{30Si}

A key step for elucidating these processes is to understand the
equilibrium structure, spectroscopy and energetics of these
molecules. For these reasons the vibrational spectroscopy of the
above molecules has been the subject of several theoretical and
experimental studies.  It is also the subject of the present paper
in which we apply a combination of an accurate {\it ab initio}
electronic structure treatment and a high-order CVPT treatment of
the vibrational problem to the CF$_4$ and SiF$_4$ molecules.

Early gas-phase spectroscopic measurements of the CF$_4$
molecule\cite{early,Jon78b} have been reviewed and supplemented by
Jones, Kennedy, and Ekberg (JKE).\cite{Jone78JCP} Earlier,
Jeannotte {\it et al.}\cite{Jea73} obtained spectra in liquid
argon solution, which as expected differ considerably from the JKE
results due to solvent shifts. Early high-resolution work on the
molecule was stimulated by the announcement\cite{Wittig} of
mid-infrared CF$_4$ lasing based on the
$\nu_2+\nu_4\rightarrow\nu_2$ transition. Esherick {\it et
al.}\cite{Eshe81JMS} determined $\nu_1$ to high resolution by
inverse Raman spectroscopy. High-resolution Raman spectroscopy of
$\nu_2$ was carried out by Lolck,\cite{Lolc81JRS} and of $2\nu_2$
and $\nu_1$ by Tabyaoui {\it et al.}\cite{Taby94JRS} The
$\nu_2+\nu_4$ combination band was measured to high accuracy by
Patterson {\it et al.}\cite{Patt80JMS} and by Poussigue {\it et
al.}\cite{Pouv2+v4} The $\nu_4$ band was studied extensively by
Jones {\it et al.},\cite{Jon78b} by Tarrago {\it et
al.},\cite{tar81} and by McDowell {\it et al.},\cite{McDo80JMS}
the latter of whom determined $\nu_4$ for all three isotopic
species with Doppler-limited resolution. Doppler-limited
measurements of the $2v_1+\nu_4$ combination band were published
by two different groups.\cite{Pine82JMS,Dang81JMS} A Fermi type 1
interaction exists between $2\nu_4$ and $\nu_3$;\cite{tak81}
detailed analyses of the $\nu_3$ and $2\nu_3$ \cite{Gaba95JMS} and
of the $\nu_3/2\nu_4$ polyad\cite{e} have been carried out.

The spectroscopic studies of CF$_4$ have been accompanied by
several theoretical (spectroscopic) 
investigations. 
Harmonic valence force fields
were first derived by by Duncan and Mills\cite{Duncan} and by
Chalmers and McKean.\cite{Chalmers} Jeannotte {\it et
al.}\cite{Jea78} fitted a harmonic potential, supplemented by
diagonal cubic and quartic stretching force constants only, to
their earlier\cite{Jea73} liquid argon solution measurements.
Brodersen\cite{Bro91} obtained a cubic force field from
experimental rotation-vibration data.
A quartic force field is said to be partly refined to
rovibrational levels below 1400 cm$^{-1},$ but detailed
information about the fitting is unavailable so
far.\cite{Lars93JMS} At higher energies, Boujut {\it et
al.}\cite{Bouj98MP} consider local and normal mode behavior in the
$\nu_3$ and $\nu_4$ ladders of tetrahedral XY$_4$ and octahedral
XY$_6$ molecules, with CF$_4$ being considered among the former.
In their analysis they employ data from \Ref{e} as
computerized into the STDS data bank.\cite{stds}  The $n\nu_3$
ladder has also been found to be a good testing ground for
comparing experiment and theory.  Using their results of
high-resolution proton energy loss spectroscopy, Maring {\it et
al.},\cite{Mari95JCP} review earlier data for $\nu_3$, $2\nu_3$,
and $3\nu_3$. They also summarize earlier unpublished data by
Heenan\cite{Heenan} who derived two sets of
Hecht\cite{Hech60JMS}-type anharmonic constants (to be denoted
Heenan I and Heenan II in the remainder of the paper) from fitting
Urey-Bradley type force fields to the Jeannotte\cite{Jea73} and
JKE data, respectively.  Maring {\it et al.} propose some
alterations based on their measurements, including an apparently
exceptionally large value for $X_{33}=-9.1$ cm$^{-1}$.

Early work on the spectroscopy of SiF$_4$ has been reviewed by
McDowell {\it et al.},\cite{McDo82JCP} who note that research on
the molecule was stimulated by the fact that the $\nu_3$
fundamental overlaps the $P$ branch of the 9400 nm band of the
CO$_2$ laser. McDowell and coworkers reported a partial set of
anharmonicity constants as well.  In recent theoretical work, the
low lying vibrational spectrum of SiF$_4$ has been
modeled\cite{Hou98CPL,Hou98AP} using an algebraic
approach.\cite{IachelloLevine}

Patterson and Pine\cite{Pat82} determined $B_0$=0.13676(3) cm$^{-1}$,
whence $r_0$=1.55982(17) \AA\cite{McDo82JCP}. In later work, Takami
and Kuze\cite{Taka83JCP} and later J\"orissen {\em et al.}\cite{Jori89CJP}
substantially revised $B_0$ upward (to 0.137780439(92) cm$^{-1}$,
consistent with a substantially shorter $r_0$=1.55404 \AA). As this paper
was being finalized, we received a preprint by Demaison {\em et al.}\cite{Demaison},
in which $r_e$=1.5524(8) is derived from a combination of {\em ab initio}
and experimental results.

In the present paper we re-investigate the spectroscopy of these
molecules using a combination of accurate {\it ab initio} anharmonic
force fields and advanced techniques for solving the vibrational
Schr\"odinger equation.  In an earlier electronic structure study,
Martin and Taylor\cite{sif4} revised the heat of vaporization of
silicon (which, among other things, is required for any {\it ab initio}
or semiempirical calculation of the heat of formation of any
silicon compound) from a benchmark {\it ab initio} calculation on
SiF$_4$ from a very precise fluorine bomb calorimetric
measurement\cite{Joh86} of the heat of formation of SiF$_4$($g$)
and a benchmark {\it ab initio} calculation\cite{sif4} of the total
atomization energy of said molecule.  In this paper, the focus is
restricted to the molecular vibrations.  In part, we were
motivated to study these molecules due to the success of a recent
study in which, an {\it ab initio} quartic force field for
methane\cite{ch4} obtained using large basis sets and coupled
cluster methods served as the starting point for several spectral
refinement studies using variational methods,\cite{carter}
low-order perturbation theory,\cite{Venu99JCP} and high-order
canonical Van Vleck perturbation theory (CVPT).\cite{Wang99JCP} In
general, we find that only the quadratic force constants and
perhaps the geometry needs to be refined, and most of the
remainder of the force field can be constrained to the {\em ab
initio} values.
\section{Ab initio anharmonic force field}

All electronic structure calculations were carried out using
MOLPRO 97\cite{molpro} running on an SGI Origin 2000 minisupercomputer
at the Weizmann Institute of Science.

Electronic correlation was treated at the CCSD(T) level
[coupled cluster with all single and double
substitutions (CCSD)\cite{Pur82} supplemented with a quasiperturbative
estimate of the contribution of connected triple
excitations\cite{Rag89}] method, using the implementation thereof in
MOLPRO.\cite{Ham92} This method is known\cite{Lee95} to yield correlation
energies that are very close to the exact basis set correlation energy
(i.e. the full configuration interaction result in the same basis set)
as long as correlation is predominantly dynamical in character. As
pointed out previously,\cite{sif4} correlation in SiF$_4$ is essentially
purely dynamical in character, and the same holds true for CF$_4$. For the
record, values for the ${\cal T}_1$ diagnostic\cite{Lee89IJQC}
(a measure for the importance of nondynamical correlation) are
only 0.0118 for SiF$_4$ and 0.0119 for CF$_4$, respectively. (In all
calculations reported here, only valence electrons were correlated.)

Geometry optimizations were carried out by
univariate polynomial interpolation.
At the reference geometry, a quartic force field in curvilinear symmetry
coordinates
was determined by recursive application of the central finite difference
formula with step size 0.01 \AA\ or radian. (Quartic contamination
was removed from the quadratic force constants.) The symmetry coordinates
are the same as those used in the previous study on CH$_4$.\cite{ch4}

The resulting curvilinear internal coordinate force field was then transformed to
rectilinear normal coordinates, then a standard second-order rovibrational
perturbation theory\cite{Papo82} was carried out by means of the
SPECTRO program.\cite{spectro,Gaw90} 
The alignment conventions for the anharmonic constants of a spherical
top follow the work of Hecht\cite{Hech60JMS} and general formulae for these
constants were taken from the paper by Hodgkinson {\it et al.}.\cite{Hodg83MP}
(See Section III.B for a discussion about the definition of the $\nu_2$ 
normal coordinates.)
The implementation in SPECTRO was slightly modified to allow
deperturbing for an arbitrary number of Fermi resonances,\cite{depert}
and a simple routine implementing Hecht's\cite{Hech60JMS} energy level
formulas up to four vibrational quanta was added.
Similar procedure has also been implemented in the curvilinear formalism,
and the relationship between the two formalism will be discussed
in Section III.B.
As in past work (e.g. \Ref{ch4}), correct alignment was verified by
slightly (0.00001 a.m.u.) perturbing the masses of two atoms,
then repeating the analysis in the asymmetric top formalism.

Given the $n^3N^4$ scaling (with $n$ the number of electrons correlated
and $N$ the number of basis functions) of CPU time for a CCSD(T) calculation,
the large number of valence electrons correlated ($n$=32) imposes certain
restrictions on the size of the basis set for a complete quartic force
field. We settled on the standard Dunning cc-pVTZ (correlation consistent
polarized valence triple zeta) basis set\cite{Dun89,Woo93} (150 contracted
Gaussian-type functions, CGTFs) for CF$_4$,
while for SiF$_4$, we opted for the cc-pVTZ+1 basis set (159 CGTFs),
where the suffix "+1" denotes the addition
of a single high-exponent $d$ function on second-row elements\cite{sio}
to accommodate inner polarization effects.
The latter has been repeatedly shown to be essential for accurate computed
properties of second-row molecules with polar bonds,\cite{sio,so2,so3} and
the addition of a mere five functions to the basis set increases CPU time
insignificantly for our purposes.
It should be pointed out that even with these relatively compact basis
sets, the evaluation of each force field took the equivalent of
six weeks of CPU time on the SGI Origin.

For purposes of validation, we in addition calculated geometries and
harmonic frequencies at the CCSD(T) level with a number of other basis
sets. For CF$_4$, we considered the (diffuse-function) augmented cc-pVTZ basis
set aug-cc-pVTZ of Kendall {\it et al.}\cite{Ken92} (230 CGTFs), while for
SiF$_4$, we considered both the standard cc-pVTZ basis set
(154 CGTFs) and the aug-cc-pVTZ+1 basis set (239 CGTFs), in which
the "+1" suffix again denotes the addition of an inner-polarization
$d$ function. The results are collected in Tables \ref{tab_adj_cf4}
and \ref{tab_adj_sif4}.

For CF$_4$, the main deficiency appears to be that the asymmetric
stretching frequency $\omega_3$ is seriously overestimated. This is
a known problem for polar bonds, particularly those involving fluorine
.\cite{Mar94CPL,Mar94JPC,Mar98CPL} Increasing the basis set size to
cc-pVQZ (275 CGTFs) or even cc-pV5Z (455 CGTFs) as a rule does not
remedy this problem\cite{Mar94CPL,Mar98CPL}
--- and given the $N^4$ scaling behavior would increase CPU time
requirements by one and two orders of magnitude, respectively, and
would therefore be essentially impossible. The use of anion
basis set (i.e. aug-cc-pV$n$Z) on at least the electronegative atoms
themselves generally leads to a dramatic improvement in basis set
convergence,\cite{Mar94CPL,Mar94JPC,Mar98CPL}
and here too we see that CCSD(T)/aug-cc-pVTZ frequencies
for CF$_4$ are much closer to the final adjusted values than
the CCSD(T)/cc-pVTZ values. While in HF, this sensitivity extends to the
anharmonicities,\cite{Mar98CPL}
in heavy-atom systems like FNO and ClNO\cite{Mar94JPC} the anharmonicities
are generally unaffected by this change. We have therefore chosen not
to calculate the entire
anharmonic force field at the CCSD(T)/aug-cc-pVTZ level,
which would have increased CPU time requirements approximately fivefold.

Interestingly, the effect of adding the anion functions is much less
pronounced for SiF$_4$ than it is for CF$_4$. The effect of adding
the inner polarization function is fairly pronounced on the bond distance
(a decrease of almost 0.01 \AA), but effects on the harmonic frequencies
are fairly minor. Here too, we judged that the CCSD(T)/cc-pVTZ+1
level represented the best compromise between accuracy and computational
cost for the entire anharmonic force field.

Purely {\it ab initio} calculation of the geometry and the harmonic part of
the potential would presumably be feasible as far as basis set convergence
and inner-shell correlation corrections are concerned, but we know from
experience\cite{Mar98CPL} that imperfections in the CCSD(T) method might
still account for (usually positive) errors of
several cm$^{-1}$ in the frequencies. Full configuration interaction
calculations  are out of the question for this system, and hence we
have followed a different track similar to previous work on silane,\cite{sih4}
and refined the `raw' CCSD(T)/cc-pVTZ and CCSD(T)/cc-pVTZ+1
force fields in the following way.

(1) Using standard second-order rovibrational perturbation theory,
the four vibrational fundamentals
for the most abundant isotopomer were obtained, together with the equilibrium
and ground-state rotational constants $B_e$ and $B_0$, respectively.
Hence (as $r_e\sqrt{B_e/B_0}$) the computed $r_0$ was obtained.

(2) Using 
the internal coordinate force field
given above, the four diagonal quadratic force constants and $r_e$ were then
iteratively adjusted such that the fundamentals and $r_0$
obtained by second-order
rovibrational perturbation theory coincide with the experimental values
to the precision given. ($F_{47}$ and all cubic and quartic force
constants were held at their {\it ab initio} computed values.)
This process converged after three iterations,
and the final force fields are given in Table \ref{tab_ff}. This force field was
then used as input for the higher-order vibrational analysis.

As can be seen in Tables \ref{tab_adj_cf4} and \ref{tab_adj_sif4},
the adjustment only entails
relatively minor changes in the geometry and harmonic frequencies.
In addition, overall agreement between the available experimental
energy levels and the values computed using standard second-order
perturbation theory is such that we
conclude that the CCSD(T)/VTZ(+1) quartic force fields are of sufficient
quality for our purposes. (Since the computed energy levels are essentially
identical to those obtained by means of second-order CVPT if no resonances
are accounted for, the reader is referred to Tables \ref{tab_cf4} and
\ref{tab_vib} and the relevant sections below for details.)

Concerning the equilibrium geometries, we note that our final $r_e$ 
values for CF$_4$ and SiF$_4$ agree with the experimentally derived
values of Brodersen\cite{Bro91} and Demaison {\em et al.},\cite{Demaison}
respectively, to within their stated uncertainties.

\section{Calculation of Vibration Energy Levels}
In this and the following two sections,
in order to assess the accuracy of the {\it ab initio} force field,
the vibrational energy levels of CF$_4$ and
SiF$_4$ are calculated using high order canonical
Van Vleck perturbation theory(CVPT) and compared to the experimental data.
The calculation is similar to that which has been recently applied to the
methane and its isotopomers. Details about this calculation can be found
in \Ref{Wang99JCP},
and only the major procedures will be summarized here.

Firstly the exact $J=0$ vibrational Hamiltonian including the
pseudo-potential $V^\prime$ term\cite{Pick72JCP} is expanded in
terms of the curvilinear normal coordinates based on the Morse
coordinates for the stretch and angle extension coordinates for
the bend. The redundancy relation between the six bond angles is
properly taken into account in the expansion. As in the case of
methane and its isotopomers,\cite{Wang99JCP} the kinetic energy
operators are expanded to fourth order, even though CVPT is
carried out to a higher order. We choose this scheme because the
kinetic contributions of order higher than four are believed to be
small, and the number of terms rapidly increases with the order of
the expansion. The quartic potential which is already in expanded
form is used throughout this work. It contributes to the CVPT up
to the second order. It is re-expressed in the coordinates
described above so that the new expansion of the potential agrees
with the original potential up to quartic terms.

The expanded Hamiltonian is separated in the form
\begin{equation}
H_v=H^{(0)} + \lambda H^{(1)} + \lambda^2 H^{(2)} + \cdots + \lambda^n
H^{(n)},
\label{ham1}
\end{equation}
where $\lambda$ is the perturbation parameter.
A succession of canonical transformations of the form
\begin{equation}
K_v = \exp \{i\lambda^n [S^{(n)},\ ]\} \cdots \exp \{i\lambda^2 [S^{(2)},\
]\}
\exp \{i\lambda [S^{(1)},\ ]\}H_v
\label{eq:trans}
\end{equation}
are then applied to the Hamiltonian of Eq. (\ref{ham1}),
where the $S^{(n)}$ are chosen such that the matrix representation of
$K_v$ is block diagonal through order $n$.\cite{Sibe88JCP,Sibe88CPC}
The blocks are characterized by one or more polyad quantum numbers
\begin{equation}
N_t = \sum_i m_i v_i,
\label{eq:const}
\end{equation}
where the integers $m_i$ define the block.  These integers
are chosen such that the resonance interactions are not perturbatively
decoupled. Details regarding the choice of the polyad quantum numbers
will be discussed later with respect to specific molecules.

Two points, regarding our analysis of this Hamiltonian are now considered
in more detail.  First, in order to obtain and analyze the eigenvalues of
the individual blocks we diagonalize each block
in a {\it symmetrized basis} to reduce the matrix size and to help assign
the calculated levels.  Second, the transformed (effective)
Hamiltonian of \Eq{eq:trans} at second
order is particularly interesting as its coefficients are
the quartic anharmonicity constants and resonance constants,
which have been conventionally used by spectroscopists to
highlight the features of the vibrational spectra.
In the following two subsections we
discuss the basis symmetrization and extraction of
spectroscopic constants as
they have not been given in our previous work.\cite{Wang99JCP}

\subsection{Symmetrization in the Cartesian basis for the $T_d$ point group}

There are generally two ways of choosing basis functions for the doubly and
triply degenerated normal modes for the $T_d$ group molecules. The first choice
(angular momentum basis) are eigenfunctions of the vibrational angular momenta.
The second choice (Cartesian basis)
is the products of harmonic oscillator functions including different Cartesian
components of the degenerate vibrations. In the angular momentum basis,
the off-diagonal terms are the spherical tensor operators defined by Hecht\cite{Hech60JMS}
and their
matrix elements are calculated using the Wigner-Eckart theorem.
However it is more difficult to treat vibrational resonances
in the angular momentum basis, and
therefore the alternative Cartesian basis is more often used.
Another advantage of the Cartesian basis is that the matrix elements are readily calculated
since they all reduce to one-dimensional harmonic oscillator matrix elements.
For both basis, the symmetrization is nonobvious and we are going to discuss
the symmetrization in the Cartesian basis below.

The Cartesian basis denoted by
$|n_1;n_{2a},n_{2b};n_{3x},n_{3y},n_{3z};n_{4x},n_{4y},n_{4z} \rangle$
will be symmetrized into the irreducible representations
of the $T_d$ point group: $A_1$, $A_2$, $E$, $F_1$ and $F_2$.
Various equivalent approaches can be used,
such as the projection operator technique,\cite{Venu99JCP,Vall99JCC}
and the promotion operator technique with
tabulated vector coupling coefficients.\cite{Halo88CPC,Halo97JCP}
The vector coupling coefficients
(also called Clebsch-Gordan coefficients)
$\langle \Gamma_{1}\Gamma_{2}ab \ket{\Gamma_{1}\Gamma_{2}\Gamma c}$
defined in
\be
\ket{\Gamma_{1}\Gamma_{2}\Gamma c}=\sum_{a,b}\ket{\Gamma_{1}\Gamma_{2}ab}
\langle \Gamma_{1}\Gamma_{2}ab \ket{\Gamma_{1}\Gamma_{2}\Gamma c}
\ee
and tabulated
in Table A20 of \Ref{Grif61} (See also Table 1 of \Ref{Tana54JPSJ})
essentially determine how the product
basis from two basis $\ket{\Gamma_{1}a}$ and
$\ket{\Gamma_{2}b}$
are further reduced according
to the irreducible representation $\ket{\Gamma c}$
of the point group.
In this work, the latter
method is adopted with some improvements
to attain unique symmetrization.

First of all, the overtone of the $E$ ($\nu_2$)
and $F_2$ ($\nu_3$ and $\nu_4$ ) normal modes are symmetrized
separately.
For the symmetrization of the overtone of the $E$ normal modes,
the promotion operators technique with tabulated vector
coupling coefficients
has been used in \Ref{Halo97JCP}.
The method is summarized as follows.
The promotion operators with ($E_a,E_b$) symmetry
are the familiar creation
operators which act on the basis by
\bea
a^\dag_a (E_a)\ket{n_a,n_b} = \sqrt{n_a+1}\ket{n_a+1,n_b}
\nonumber \\
a^\dag_b (E_b)\ket{n_a,n_b} = \sqrt{n_b+1} \ket{n_a,n_b+1}.
\eea
Taking the promotion operators to belong to
($E_a,E_b$) symmetries, the symmetry of the
resultant promoted basis function can be identified according
to the vector coupling coefficients.
Starting from the $v$=1 basis function, one
can derive all the symmetrized basis functions
until $v$ reaches a given number similar to climbing
a $v$ ladder.
A computer code is desirable if one wants to symmetrize
a high overtone. One problem is that
spurious states with the same symmetry may arise if the promotion
operator acts on every basis, and they are not orthogonal.
For example, with promotion operators acting on
the full $3\nu_2$ symmetrized basis ($A_1+A_2+E$), three
nonorthogonal
$E$ states are obtained while only two are allowed
for $4\nu_2$.
These spurious states are removed by using
the Gram-Schmidt orthogonalization
technique.
The final orthogonal symmetrized basis is not unique,
but all possible solutions lead to symmetrized Hamiltonian matrix blocks.

An alternative method, which yields unique
symmetrization coefficients, is obtained
by finding the exact decomposition of the angular momentum basis
function into the Cartesian basis wavefunction.
 Once this is done,
the symmetrization is trivial as the angular momentum basis
function is readily symmetrized according to the value of $l$ modulo 3.
This task seems quite complicated since the
angular momentum basis functions are basically
the associated Laguerre polynomials and the
Cartesian basis functions are essentially products
of two Hermite polynomials.\cite{Paul35,Shaf44RMP}
In fact, this
complication can be circumvented by using
the circular promotion operators defined by \cite{Cohe77}
\bea
a_d^\dag = {1 \over \sqrt{2} }(a_a^\dag + i a_b^\dag)
\nonumber \\
a_g^\dag = {1 \over \sqrt{2} }(a_a^\dag - i a_b^\dag)
\eea
They act on the angular momentum basis by
\bea
a_d^\dag \ket{v^l} = \sqrt{{(v+l) \over 2} +1} \ket{(v+1)^{l+1}}
\nonumber \\
a_g^\dag \ket{v^l} = \sqrt{{(v-l) \over 2} +1} \ket{(v+1)^{l-1}}
\eea
and on the Cartesian basis by
\bea
a_d^\dag \ket{n_a,n_b} = \sqrt{n_a + 1 \over 2} \ket{n_a+1,n_b}
                        + i \sqrt{n_b + 1 \over 2} \ket{n_a,n_b+1}
\nonumber \\
a_g^\dag \ket{n_a,n_b} = \sqrt{n_a + 1 \over 2} \ket{n_a+1,n_b}
                        - i \sqrt{n_b + 1 \over 2} \ket{n_a,n_b+1}
\eea
With the circular promotion operators
acting on the $v=0$ ground state $\ket{0^0}=\ket{0,0}$
in the two basis sets, respectively, one obtains the relationship for the $v=1$
basis
\bea
\ket{1^{\pm 1}}={\sqrt{2} \over 2} (\ket{10} \pm i \ket{01})
\label{Eab}
\eea
Repeating this operation up to a given quantum number $v$ leads
to relationship of the two basis up to $v$.
The next step is to transform the angular momentum
basis from the $(E^+,E^-)$ to
$(E_a,E_b)$ symmetries. It is well known
that $\ket{v^{\pm l}}$ ($l=3p + 1$, with $p$ being any integer)
form a basis for $E^{\pm}$ irreducible representations.
Following a convention which is learned from \Eq{Eab}, the
$(E_a,E_b)$ symmetry basis functions are defined by
\bea
\ket{v^{|l|};E_a} &=&
{\sqrt{2} \over 2} (\ket{v^{+l}} + \ket{v^{-l}}) = \sqrt{2} \, {\rm Re}(\ket{v^{+l}})
\nonumber \\
\ket{v^{|l|};E_b} &=&
{\sqrt{2} \over 2i} (\ket{v^{+l}} - \ket{v^{-l}}) = \sqrt{2} \, {\rm Im}(\ket{v^{+l}})
\label{Eab1}
\eea
where $l=3p+1$. The $A_1/A_2$ symmetry basis functions are defined similarly by
\bea
\ket{v^{|l|};A_1} &=&
{\sqrt{2} \over 2} (\ket{v^{+|l|}} + \ket{v^{-|l|}}) = \sqrt{2} \, {\rm Re}(\ket{v^{+|l|}})
\nonumber \\
\ket{v^{|l|};A_2} &=&
{\sqrt{2} \over 2i} (\ket{v^{+|l|}} - \ket{v^{-|l|}}) = \sqrt{2} \, {\rm Im}(\ket{v^{+|l|}})
\label{Eab2}
\eea
where $l=3p $. There is only $A_1$ symmetry if $l$ is zero.
Pak {\it et al.} \cite{Pak97JCP} have also discussed a similar scheme to
transform the angular momentum basis functions to the
real cosine and sine form, which are members of 
the irreducible representation 
of the $D_{3h}$ group.
It should be noted that the above results can be applied to
other analogous doubly degenerated modes such as the $E$ mode in
the $C_{3v}$ group
and the $E_g$ mode in the $O_{h}$ group.
The above procedure can be easily implemented by a computer code.
Results up to $v=8$ are listed in Table \ref{e_over}, which are
obtained from a code in the Maple\cite{Maple} computer algebra language.
These results may be useful for other applications.
A fortran code is also written for the same purpose and is to be
combined with the symmetrization of the $F_2$ normal mode to form
a general symmetrization package.

Now we turn to the
symmetrization of the overtone of the $F_2$ normal mode.
This can also be achieved using the
promotion operators
together with the vector coupling coefficients.
However there exist general rules on how a given
overtone is symmetrized without resorting to
promotion of lower states.
These general rules can
be derived using the projection operator methods.
Roughly speaking, one obtains $AE$ symmetry when the
three Cartesian quantum numbers are all even or odd,
and $F$ symmetry when they are mixed even and odd.
Detailed symmetrization coefficients have been given
in Table 2 of \Ref{Bouj98MP}
for the overtone of the triply degenerated modes
($F_1$ and $F_2$ modes of $T_d$ group)
of $O_h$ and $T_d$ group.

Finally given the fact that the $E$ overtone and
$F_2$ overtone have been symmetrized separately,
the combination bands in the form ($n_1n_2n_3n_4$)
can be symmetrized using
the vector coupling coefficients, as has been
demonstrated in \Ref{Halo97JCP}. In this process,
no spurious states arise as the
basis functions joining the product
are from different normal modes.
All the above steps have been implemented as
a Fortran code which
writes to a file the symmetrization
coefficients for all the combination bands in a 
polyad with a given polyad number.  This is highly
desirable since one may encounter completely different
polyad numbers such as those for CH$_4$, CF$_4$
and SiF$_4$. These symmetrization coefficients
are further used in constructing the polyad block matrix of
the transformed Hamiltonian.

It is interesting to compare with the 
basis symmetrization for the $T_d$ group in the 
algebraic model.\cite{Lemu94JCP,Ma96PRA,Fran99JMS} A key difference 
is that the basis functions symmetrized there are products
of ten functions based on the ten internal
coordinates. 
Due to the redundancy between the six bond-angle coordinates
considerable effort was required to remove the
so-called spurious states.\cite{Lemu94JCP,Ma96PRA}
Their symmetrization procedure is basically the same
as the promotion operator technique with
tabulated vector coupling coefficients.\cite{Halo88CPC,Halo97JCP}
Lemus and Frank\cite{Lemu94JCP} have also tabulated and derived 
part of the vector coupling coefficients in the $T_d$ group,
unaware of an earlier full tabulation.\cite{Tana54JPSJ,Grif61}

\subsection{Extraction of spectroscopic constants}

The spectroscopic constants we want to
extract from our second order transformed Hamiltonian
are specifically the quartic anharmonicity constants $X_{ij}$,
$G_{ij}$, the tensor constants $T_{ij}$ and $S_{ij}$ and
the cubic $K_{i,jk}$ and quartic resonance
constants $K_{i,jkl}$ and $K_{ij,kl}$.
The quartic anharmonicity constants $X_{ij}$ and $G_{ij}$
have their usual definitions. The tensor constants
were defined by Hecht\cite{Hech60JMS} as coefficients of spherical
tensor operators which cause splittings of the same symmetry
components of the overtone and combination bands of spherical
top molecules.  This procedure has been used in earlier
work on the acetylene molecule,\cite{Sibe96JCP} however the
procedure used was not described. More importantly, the
procedure has not been applied to molecules of $T_d$ symmetry which
present a special challenge.  For these reasons we describe the procedure
in full detail here.

The apparent difficulty of the extraction is that
our CVPT results for the transformed Hamiltonian
are expressed as expansions of
creation-annihilation operators in normal form\cite{Sibe88JCP,Sibe88CPC}
--- the creation operators are put before the annihilation operators.
These expansion terms need to be rearranged to obtain
the traditional spectroscopic Hamiltonian.
A simple example based on second order perturbation theory
serves to illustrate this point.
At second order the normal form expansion includes contributions, such as
$(a_1^\dag)^2 (a_1)^2={\hat n}_1({\hat n}_1-1)$,
$a_1^\dag a_1= \hat{n}_1$ plus a constant.
It is found that by rearranging the normal form terms
as powers of $(\hat{n_1}+{1\over 2}),$ all the linear contributions
to second order of perturbation theory {\it cancel}, leaving the Hamiltonian
in the standard spectroscopic form where only quadratic terms
such as $(\hat{n_1}+{1\over 2})^2$ contribute.
With either choice of expansion the leading term will have the same coefficient
which is $X_{11}.$

A similar problem has been studied
by Hodgkinson {\it et al.}\cite{Hodg83MP} who have suggested the same
idea of extracting spectroscopic constants
from the second order transformed Hamiltonian, although
the Hamiltonian discussed there is in rectilinear normal
coordinates and the creation-annihilation operators
are not rearranged to normal form.
Furthermore, the extraction of constants related with doubly and triply degenerated
modes requires one more step than the above simple example of $X_{11}$.
For example, the tensor constants
$T_{23}$ and anharmonicity constants
$X_{23}$ are determined simultaneously by
\bea
{\rm Coef}(\hat{n}_{2a}\hat{n}_{2z}) = X_{23} + 8 T_{23}
\nonumber \\
{\rm Coef}(\hat{n}_{2b}\hat{n}_{2z}) = X_{23} - 8 T_{23}
\label{eq:T23}
\eea
where Coef$(\hat{o})$ denotes coefficient of operator $\hat{o}$,
and $\hat{n}_{2a}\hat{n}_{2z}$ and $\hat{n}_{2b}\hat{n}_{2z}$
are the leading terms in the normal form Hamiltonian.
These formulae are readily obtained by rewriting the $O_{23}$ operator
which was defined by Hecht [Eq. (8) of \Ref{Hech60JMS}] as an operator
whose coefficients are $T_{23}$.
Such formulae for all the anharmonicity and tensor constants
have been given in Table 5 of \Ref{Hodg83MP}.

Here one has to take care of a convention problem regarding the definition of normal/symmetry
coordinates for the $\nu_2$ ($E$) mode of a tetrahedral molecule.
We notice that Jahn defined the symmetry
coordinate $(Q_{2a},Q_{2b})$ on page 472 of \Ref{Jahn38PRS1}
(See Fig. 1 of \Ref{Jahn35APL} for the numbering of the
four peripheral atoms). His definition and the vector coupling coefficients
for the $E$ mode coupling other modes
(Table II of \Ref{Jahn38PRS2}) were adopted by Hecht in \Ref{Hech60JMS}.
Unfortunately the symmetry coordinates $(S_{2a},S_{2b})$
for the $\nu_2$ mode we and many other researchers\cite{Gray79MP,ch4,Halo97JCP,Venu99JCP}
adopted are those of Mills.\cite{Mill58MP,Gray79MP}
After some manipulations, it is found that the two definitions are related such that
$Q_{2a}=-S_{2b}, Q_{2b}=S_{2a}$. If one uses Mills' definition, all the Hecht formulae
regarding the $\nu_2$ mode should be modified according to the above relationship.
The only significant change occurs with the $O_{23}$ operator for which a minus
sign needs to be attached to Hecht's definition. Only after this
correction is considered, can the Table 5 of Hodgkinson {\it et al.} be
reconciled with Hecht's definition. \Eq{eq:T23} is a pertinent example.
Finally, it is noted that the vector coupling coefficients tabulated by
Tababe and Sugano\cite{Tana54JPSJ} and Griffith\cite{Grif61}
agree with Mills' definition of the
symmetry coordinates of $\nu_2$.

The extraction of the resonance constants is straightforward
in the creation-annihilation form since they are usually defined
in this form.
One just needs to take care of constructing
the correctly symmetrized Hamiltonian terms
for the resonance.
To this end, several equivalent approaches
can be used.
Other than the method by Hecht \cite{Hech60JMS,Robi76MP} for
constructing the correctly symmetrized potential tensor operators,
we choose to follow an elegant and more general approach by
Halonen\cite{Halo97JCP} who used the tabulated vector
coupling coefficients
to construct the correctly symmetrized
Hamiltonian terms from any combination of operators
such as coordinates, momenta and creation-annihilation
operators.
Methane has a variety of cubic and quartic resonances
due to its approximate 2:1 ratio of the stretch
and bend frequency. Halonen has given for the first time
all the correctly symmetrized
Darling-Dennison type quartic resonance terms
in \Ref{Halo97JCP} using the vector coupling methods.
The resonance constants encountered in CF$_4$ and SiF$_4$
are limited compared to methane.
Up to second order resonance, they are
(See also Eqs. (\ref{res2a}) -- (\ref{res2c}) and (\ref{res1a}) -- (\ref{res1c}))
\bea
H/hc &=& K_{1,22} \left[ a_1^\dag (a_{2a}^2 + a_{2b}^2) +{\rm h.c.} \right]
\\
H/hc &=& K_{3,44} \left[a_{3x}^\dag a_{4y} a_{4z} + a_{3y}^\dag a_{4z} a_{4x}
                  +a_{3z}^\dag a_{4x} a_{4y} +{\rm h.c.} \right]
\eea
for CF$_4$ and
\bea
H/hc &=& K_{1,44} \left[ a_1^\dag (a_{4x}^2 + a_{4y}^2 + a_{4x}^2) +{\rm h.c.} \right]
\\
H/hc &=& K_{1,222} \left[ a_1^\dag a_{2a}(a_{2a}^2 - 3 a_{2b}^2) +{\rm h.c.} \right]
\label{k1222}
\eea
for SiF$_4$. The quartic resonance with coefficient $K_{1,222}$ is of a new type.
We shall show how it is derived to illustrate the vector coupling method.
The goal is to construct the totally symmetric ($A_1$ symmetry) quartic terms
from one creation operator ($a_1^\dag$) with $A_1$ symmetry and two annihilation operators
$(a_{2a},a_{2b})$ with $E$ symmetry. Apparently there is only one such term.
First the $E_a/E_b$ symmetry quadratic terms are constructed from
products of $(a_{2a},a_{2b})$ and themselves
as follows
\bea
E_a \: &:&\: {1 \over \sqrt{2}} (-a_{2a}^2 + a_{2b}^2)
\nonumber \\
E_b \: &:&\: \sqrt{2} a_{2a} a_{2b}
\label{k1222a}
\eea
Then the $A_1$ symmetry cubic terms are constructed from
products of $(a_{2a},a_{2b})$
and the two terms of \Eq{k1222a} as follows
\bea
A_1 \: &:&\: {1 \over \sqrt{2}} \left[ {1 \over \sqrt{2}} (-a_{2a}^2 + a_{2b}^2) \cdot a_{2a}
 + \sqrt{2} a_{2a} a_{2b} \cdot a_{2b} \right]
\nonumber \\
&&\:= -{1 \over 2} a_{2a}(a_{2a}^2 - 3 a_{2b}^2)
\label{k1222b}
\eea
Finally the $A_1$ symmetry quartic term is obtained as product of $a_1^\dag$ and terms of
\Eq{k1222b} plus the Hermite conjugate of the product.
The coefficient of the final resonance term in \Eq{k1222} has been redefined for convenience.

An interesting point is that the anharmonicity and tensor
constants extracted using the method above agree exactly with
standard second-order perturbation theory in rectilinear normal
coordinates as implemented in SPECTRO, provided that
the same force field is used and no resonances are considered. 
This seems to be a surprise, since
the reported CVPT results are based on curvilinear coordinates.
McCoy and Sibert\cite{McCo92MP} have explained this as follows.
When using dimensionless normal coordinates the perturbation
parameter in the expansion of Eq. (1) can be taken as
$\hbar^{1/2}$.  Here terms of order $n$ in Eq. (1) are of order
$\hbar^{(2+n)/2}.$  This is true regardless of using rectilinear
or curvilinear normal coordinates.  Since both sets have identical
zero-order Hamiltonians, the energies have to be identical order
by order. If one uses perturbation theory to transform to
block-diagonal Hamiltonian, which is subsequently diagonalized,
the above argument breaks down, and notable differences have been
found.\cite{McCo91JCP1,McCo91JCP2} In this work, a multi-resonance
Hamiltonian has been considered for CF$_4$ and SiF$_4$. Therefore
the spectroscopic constants are extracted from the second order
CVPT transformed Hamiltonian, and some of them are different
from those obtained using SPECTRO with resonances deperturbed.
More details regarding this point are given with respect to
CF$_4$ and SiF$_4$ molecules in the following section.

\section{Results for CF$_4$}

The choice of the polyad quantum numbers
for CF$_4$ is based on the following considerations.
The harmonic frequencies
as calculated from the {\it ab initio} force fields
($\omega_1$ = 921.57 cm$^{-1}$, $\omega_2$ = 439.91 cm$^{-1}$,
$\omega_3$ = 1303.01 cm$^{-1}$, $\omega_4$ = 637.89 cm$^{-1}$)
suggest that three independent resonances
are possible:
\bea
\omega_1  & \approx & 2 \omega_2
\label{res2a}
\\
\omega_3  & \approx & 2 \omega_4
\label{res2b}
\\
3\omega_2  & \approx & \omega_3
\label{res2c}
\eea
The frequency differences between the above pairs are only 42 cm$^{-1}$,
27 cm$^{-1}$ and 17 cm$^{-1}$, respectively. Since these small differences will
go into the denominator of the expression of $S^{(n)}$, they will
possibly lead to divergence of the perturbation theory when the
coupling strength between the two pair states are not negligible.
In practice we found that all the three resonances ought to be
considered by keeping them in the $K_v$ to ensure good convergence when
the perturbation theory is carried up to the sixth order. It should
be noted that the resonance in Eq. (\ref{res2c}) is actually
a fourth order resonance,
because the symmetry is unmatched between
$3\nu_2$ and $\nu_3$ at second order.
Due to the three independent resonances, only one good
quantum number remains, which should be
orthogonal to the three resonance vectors $(1,-2,0,0)$, $(0,0,1,-2)$ and
$(0,-3,1,0)$. It is then easy to find the remaining good quantum
number to be
\be
N=4v_1 + 2v_2 + 6v_3 + 3v_4,
\ee
which will be used in the perturbation calculations reported below.

In Table \ref{tab_cf4}, we collected all the reliable experimental band origins
of the three isotopes. There are 14 high resolution data (with resolution
better than 0.01 cm$^{-1}$). All the other data with typical 0.1 and 1 cm$^{-1}$
resolution are from the work of Jones {\it et al.}.\cite{Jone78JCP}
There are some other low-resolution data compiled in \Ref{Jone78JCP}, which
we found to be unreliable and were discarded. The calculated energy levels based on the
adjusted {\it ab initio} force field of Table \ref{tab_ff} agree well with the observations.
We then further refined all five quadratic force constants to the observed
levels with uncertainty for each level determined by the experimental
precision, using a fast convergent second order least squares method
where the Hessian is approximated as products of first derivatives.\cite{Mard98JMS}
The refined constants are given in Table \ref{cf4_cst} and
the comparison of experimental and calculated energy levels is given in Table \ref{tab_cf4}.
The RMSD (root mean
squares deviations) based on the refined potential
is 0.179, 0.083 and 0.083 cm$^{-1}$ at second, fourth
and sixth order CVPT, respectively, whereas the
corresponding RMSD based on the {\it ab initio} potential in Table \ref{tab_ff}
is 0.095, 0.190 and 0.186 cm$^{-1}$.
This shows that the {\it ab initio} potential is a rather good
as an initial potential for the fitting.
The following remarks can be made by examining the calculated results
in Table \ref{tab_cf4}.

(1) The CVPT calculation converges very well. The energy level changes typically
on the order of 0.1 cm$^{-1}$ by comparing the fourth and sixth order results.
This contrasts with the case of CH$_4$
where the bending overtone (4$\nu_4$)
varies by about three wavenumbers at the eighth order.\cite{Wang99JCP}
Therefore the fourth order CVPT is used for the fitting.
The contribution due to the $V^\prime$ terms is small. For example,
the largest contribution to the fundamentals
is 0.057 cm$^{-1}$ for (0001;$F_2$) band of $^{12}$CF$_4$ using sixth order
CVPT.

(2) For $^{12}$CF$_4$,
the resonance $\omega_3  \approx 2 \omega_4$ is the most pronounced.
Even the (0010;$F_2$) fundamental has 7 \% admixture from (0002;$F_2$).
The resonance
$\omega_1 \approx 2\omega_2$ is less pronounced and the resonance
$3\omega_2 \approx \omega_3$ can be ignored. The above remarks also
apply for the $^{13}$CF$_4$ whereas all the three resonances seem
to be less pronounced for $^{14}$CF$_4$.

The pronounced resonance between $\omega_3  \approx 2 \omega_4$
means that previous treatment
for the $\nu_3$ ladder states where only
the $\nu_3$ overtones are considered are inadequate
.\cite{Gaba95JMS,Bouj98MP,Mari95JCP}

(3) The calculations for the three (0020) bands using either the
{\it ab initio} or fitted force field cannot reproduce well
the observations which were obtained through a rovibrational analysis
of the isolated (0020) bands.\cite{Gaba95JMS} We cannot fully understand these
discrepancies. But it is clear that the (0020) band is
in Fermi resonance with the (0012) and (0004) bands, although
the major perturbant (0012) is a little farther away.
For example, the three (0012;$F_2$) bands are respectively at
2536.50 cm$^{-1}$, 2541.66 cm$^{-1}$, and 2544.45 cm$^{-1}$.
We have also noted that near the
experimental (0020;$F_2$) band origins (2561.91 cm$^{-1}$), there is a band
calculated to be 2561.04 cm$^{-1}$ with 82\%(0302;$F_2$) and 14\%(0310;$F_2$).
However it is very unlikely that this band should be as strong as
the (0020;$F_2$) band. Further theoretical and experimental work on
this band system appears to be desirable.

Finally, the vibrational spectroscopic constants for all the isotopes
of CF$_4$ and SiF$_4$ obtained from the second order transformed CVPT Hamiltonian
(curvilinear formalism)
are given in Table \ref{tab_tensor}. With these constants 
one can reproduce the second order CVPT calculation by constructing
the block diagonal Hamilonian matrices. 
The corresponding constants in the rectilinear formalism, where different,
are given in footnotes to said table. The differences (for the Fermi
resonance constants and for such anharmonic constants as are affected by
the Fermi interactions) illustrate the points made in the previous section.

Discrepancies between the presently computed anharmonicity constants
and the empirically derived sets of Heenan\cite{Heenan} mostly reflect
the limitations of the latter in terms of available data. There is no
indication for the very high $X_{33}=-9.1$ cm$^{-1}$ as
suggested by Maring {\em et al.}\cite{Mari95JCP} We also note that both
Heenan sets of harmonic frequencies differ quite substantially from our
best values, and are confident that the latter are more reliable.

\section{Results for SiF$_4$}

The choice of the polyad quantum numbers
for SiF$_4$ is based on the following considerations.
 The harmonic frequencies
as calculated from the {\it ab initio} force fields
($\omega_1$ = 806.3 cm$^{-1}$, $\omega_2$ = 265.2 cm$^{-1}$,
$\omega_3$ = 1044.0 cm$^{-1}$, $\omega_4$ = 389.3 cm$^{-1}$)
suggest that three independent resonances
are possible:
\bea
\omega_1  & \approx & 2 \omega_4
\label{res1a}
\\
\omega_1  + \omega_2 & \approx &  \omega_3
\label{res1b}
\\
\omega_1  & \approx & 3 \omega_2
\label{res1c}
\eea
The differences between the above pairs are only 27 cm$^{-1}$,
27 cm$^{-1}$, and 11 cm$^{-1}$, respectively. Again all three resonances ought to be
considered to ensure good convergence when
the perturbation theory is carried up to sixth order. It should
be noted that the resonance in Eq. (\ref{res1b})
is actually a third order resonance,
because the symmetry is unmatched
between $\nu_1$ + $\nu_2$ ($F_2$) and $\nu_3$ ($E$)
at first order.
Following the approach of the previous section the good quantum
number, to be used in the perturbative calculations, is found to be
\be
N=6v_1 + 2v_2 + 8v_3 + 3v_4
\ee
which we are going to use in the perturbation calculations reported below.

The sixteen observed band centers from \Ref{McDo82JCP} and the values
of the second, fourth and sixth order CVPT
calculations are given in Table \ref{tab_vib}.
The following remarks can be made by examining Table \ref{tab_vib}.

(1) The calculated values converges to a few hundredths of a wavenumber
by comparing the fourth and sixth order results.
The convergence is better than in the case of CF$_4$.
The contribution due to the $V^\prime$ terms is small. For example,
the largest contribution to the fundamentals
is 0.0016 cm$^{-1}$ for (0001;$F_2$) band of $^{28}$SiF$_4$ using sixth order
CVPT. This is smaller than that of the $^{12}$CF$_4$
since SiF$_4$ is heavier.

(2) The agreement between
calculation and experiment is rather good with a RMSD
of 0.59, 0.74 and 0.73 cm$^{-1}$ at second, fourth
and sixth order CVPT, respectively.
It should be noted that the slightly better agreement at the second order is fortuitous
as the second order results still do not converge.
On the whole, the CVPT results are remarkable in spite of the fact that
the CVPT calculation uses the {\it ab initio} force field where
only four diagonal quadratic force constants were optimized
to the four experimental fundamentals. The success of the perturbation
calculation is mainly due to the small anharmonicity for this relatively
heavy-atom molecule (cfr. the $\omega_i-\nu_i$ values in Tables \ref{tab_adj_cf4}
and \ref{tab_adj_sif4}).

(3) In Table \ref{tab_vib} our calculations are compared with those of the $U$(2)
algebraic model.\cite{Hou98CPL} Although the $U$(2) algebraic model yields
a similar RMSD (0.79 cm$^{-1}$),
nine parameters were employed to fit the sixteen vibrational
data. A fit of similar quality is also reported by the same authors\cite{Hou98AP} using a boson realization model. 
However, in both models the polyad number $N=v_1+v_2+v_3+v_4$ is used and
the Fermi resonance $\omega_1 \approx 2\omega_4$ is considered in neither.

(4) It can be seen from Table \ref{tab_vib} that the resonance $\omega_1  \approx 2 \omega_4$
is most prominent. The other two resonances are also considered to ensure
good convergence at sixth order.
To neglect these two resonances will considerably reduce the size of the matrix associated with
a given value of the quantum number, so
that high overtone states can be relatively easier to calculate.
The lack of strong resonance in $\nu_3$ and its overtone ensures that the isolated treatment
of the $\nu_3$ overtone is a good approximation.\cite{Pat82}
Patterson and Pine has done such a calculation, and find the Cartesian basis is much inferior
to the angular momentum basis due to the near degeneracy of the Cartesian basis. This near
degeneracy is specific to SiF$_4$ since the relation $G_{33}=8T_{33}$
is nearly satisfied.\cite{Pat82} However, as far as the calculation of
the vibrational energy is concerned, both basis sets can be used.

The partial set of empirically derived anharmonicity constants 
(Table \ref{tab_tensor}) of McDowell
{\em et al.}\cite{McDo82JCP} agrees fairly well with our computed values,
while their derived harmonic frequencies agree to within their stated
uncertainty with our own best values.

\section{Conclusions}

The anharmonic vibrational spectra of the all-heavy atom spherical tops
CF$_4$ and SiF$_4$ have been treated by the combination of an accurate 
{\em ab initio} anharmonic force field and high-order canonical Van
Vleck 
perturbation theory (CVPT).

The anharmonic part of the potential energy surface is evidently very
well described at the CCSD(T)/cc-pVTZ level for CF$_4$, and at the 
CCSD(T)/cc-pVTZ+1 level for SiF$_4$, where the `+1' notation refers
to the addition of a high-exponent $d$ function on second-row atoms. 
The harmonic frequencies of CF$_4$ exhibit substantial errors at the 
CCSD(T)/cc-pVTZ level, which disappear upon addition of diffuse
functions 
to the basis set. 

The force field is subsequently slightly refined by adjusting the 
equilibrium geometry and the diagonal quadratic force constants
(in this case, five parameters) such that a standard second-order
rovibrational perturbation theoretical treatment reproduces the
experimental fundamentals and ground-state rotational constants of
the molecules. The adjustments involved are fairly minor. 

These force fields were then used as input for CVPT calculations
up to sixth order inclusive. While agreement with experiment is
fortuitously quite good at second order in CVPT, consistent convergence to
0.1 cm$^{-1}$ in the energy level is only achieved at sixth order.
However, the truncation error at second order is much less significant
than in the case of hydride systems like CH$_4$ and SiH$_4$.

RMS deviation (RMSD) between computed (sixth order CVPT) and observed
energy levels 
with the adjusted ab initio potentials is 0.19 cm$^{-1}$ for CF$_4$ and 0.73
cm$^{-1}$ for SiF$_4$, the latter to some extent reflecting the lesser 
accuracy of the experimental data. 
In the case of CF$_4$, improvement with experiment could be somewhat
further improved (to RMSD=0.08 cm$^{-1}$) by re-adjusting both diagonal
and 
off-diagonal quadratic
force constants to the complete set of experimental vibrational level
information. Experimentation with refinement of additional force field
parameters yielded no further improvement. 

For CF$_4$, three resonances, Eqs. (\ref{res1a})--(\ref{res1c}), 
were considered,
of which only one ($\omega_3 \approx 2 \omega_4$) is important.
This shows that an isolated $\nu_3$ overtone ladder model is inadequate.
For SiF$_4$, three resonances, Eqs. (\ref{res2a})--(\ref{res2c}), 
were considered,
of which only one ($\omega_1 \approx 2\omega_4$) is important.

An improved approach for symmetrizing combination bands in the
Cartesian basis for the $T_d$ group is proposed.
We also demonstrate how anharmonic spectroscopic constants
$X_{ij},G_{ij},T_{ij},S_{ij}$ can be extracted from the second-order
CVPT transformed Hamiltonian (in curvilinear internal coordinates) 
for $T_d$ molecules. In the absence of
resonances, the results are
identical to those obtained by standard second-order perturbation theory
in rectilinear normal coordinates. Differences occur when the constants
are being deperturbed for Fermi resonances. 
Accurate sets of quartic spectroscopic constants
for the isotomomers of CF$_4$ and SiF$_4$ are obtained. Agreement with 
previously published empirically derived sets of anharmonicity constants
is fairly good for SiF$_4$, but less satisfactory for CF$_4$. 

In order to stimulate further research on these molecules, sixth-order
CVPT energy level
predictions up to polyad number $N=24$ for $^{\{12,13,14\}}$CF$_4$ and
$^{\{28,29,30\}}$SiF$_4$ have been made available on the World Wide Web
at the Uniform Resource Locator (URL)
\url{http://theochem.weizmann.ac.il/web/papers/cf4sif4.html}. The force
fields themselves are available in machine-readable form at the same URL.

\section*{Acknowledgments}

Prof. M. S. Child is thanked for helpful discussions on
the basis set symmetrization.
JM is a Yigal Allon Fellow and the incumbent of the Helen and Milton
A. Kimmelman Career Development Chair, acknowledges support from
the Minerva Foundation, Munich, Germany, and would like to thank
Dr. Timothy J. Lee (NASA Ames Research Center) for helpful discussions
and Prof. Jean Demaison (Universit\'e de Lille I) for a preprint of
\Ref{Demaison}.
XGW thanks the Royal Society KC Wong Fellowship for support.

\baselineskip=18pt

\begin{table}
\caption{Computed (CCSD(T)) and observed bond distances (\AA), harmonic and
fundamental frequencies (cm$^{-1}$)\label{tab_adj_cf4} for $^{12}$CF$_4$}
{\small
\begin{tabular}{lccccccccc}
&\multicolumn{3}{c}{cc-pVTZ unadj.} &aug-cc-pVTZ&
\multicolumn{3}{c}{cc-pVTZ adjusted} & \multicolumn{2}{c}{Expt.\cite{McDo82JCP}}\\
 & $r_e$ & $r_0$ &$r_z$&$r_e$ &$r_e$& $r_0$ &$r_z$&$r_e$&$r_0$\\
\hline
&1.31919  &  1.32310  & 1.32389& 1.32112  & 1.31526 & 1.31925 &1.32004&1.3151(17)$^a$&  1.319247(1)$^b$ \\
\hline
$i$ & $\omega_i$ & $\nu_i$ & $\omega_i-\nu_i$ & $\omega_i$ &
$\omega_i$ & $\nu_i$ & $\omega_i-\nu_i$ & $\omega_i$ & $\nu_i$ \\
\hline
1&922.80   & 910.75    & 12.06 &   915.2  & 921.57  & 909.07  & 12.50  &&  909.0720(1) \cite{Eshe81JMS}\\
2&440.00   & 435.59    & 4.40  &   435.2  & 439.91  & 435.40  & 4.51   &&  435.399(10) \cite{Lolc81JRS}\\
3&1322.25  & 1303.15   & 19.10 &  1301.3  & 1303.01 & 1283.66 & 19.35  && 1283.66429(12)\cite{e} \\
4&638.81   & 632.22    & 6.59  &   630.4  & 637.89  & 631.06  & 6.84   && 631.05890(13) \cite{e}\\
\end{tabular}
}

$^a$ \Ref{Bro91}.

$^b$ From $B_0$=0.19118709(32) cm$^{-1}$.\cite{Gaba95JMS}

\end{table}

\begin{table}
\caption{Computed (CCSD(T)) and observed bond distances (\AA), harmonic and
fundamental frequencies (cm$^{-1}$)\label{tab_adj_sif4} for $^{28}$SiF$_4$}
{\small
\begin{tabular}{lcccccccccc}
&cc-&aug-cc-\\
&pVTZ&pVTZ+1&\multicolumn{3}{c}{cc-pVTZ+1 unadj.} &
\multicolumn{3}{c}{cc-pVTZ+1 adjusted} & \multicolumn{2}{c}{Expt.\cite{McDo82JCP}}\\
 & $r_e$ &$r_e$ &$r_e$ & $r_0$ &$r_z$&$r_e$ & $r_0$ &$r_z$&$r_e$&$r_0$\\
\hline
&1.56949&1.56332&1.56134  &  1.56368  & 1.56453 & 1.55182 & 1.55404 &1.55489&1.5524(8)$^a$&  1.55404$^b$ \\
\hline
$i$ & $\omega_i$ &$\omega_i$ &$\omega_i$ & $\nu_i$ & $\omega_i-\nu_i$ &
$\omega_i$ & $\nu_i$ & $\omega_i-\nu_i$ & $\omega_i$ & $\nu_i$ \\
\hline
1&  794.9 & 794.1  &797.86   & 792.19    & 5.67     & 806.10  & 800.60  & 5.50  &  807.1(12) &  800.6\\
2&  259.8 & 258.8  &263.18   & 262.13    & 1.05     & 265.20  & 264.20  & 1.00  &  267(3)    &264.2\\
3& 1036.5 & 1029.8 &1037.49  & 1024.31   & 13.18    & 1044.04 & 1031.40 & 12.64 & 1044.2(12) &1031.3968 \\
4&  384.3 &  382.7 &387.61   & 386.67    & 0.94     & 389.31  & 388.44  & 0.87  & 389.8(9) &388.4448\\
\end{tabular}
}

$^a$ \Ref{Demaison}.

$^b$ From \Ref{Jori89CJP}, $B_0$=0.137780439(92) cm$^{-1}$. In older work\cite{McDo82JCP,Pat82},
$B_0$=0.13676(3) cm$^{-1}$ and hence $r_0$=1.55982(17) \AA.

\end{table}

\begin{table}
\caption{Quadratic, cubic and quartic force
constants (aJ/\AA$^m$radian$^n$) for SiF$_4$ and CF$_4$\label{tab_ff}}
\begin{tabular}{lrr|lrr|lrr}
& SiF$_4$ & CF$_4$ & & SiF$_4$ & CF$_4$ & & SiF$_4$ & CF$_4$ \\
\hline
$F_{11}$ &     7.27355   &  9.50679  &   $F_{22}$   &   0.63194   &   1.24913  &   $F_{44}$ &    6.40971 &   6.13519  \\
$F_{74}$ &   --0.34328   &--1.03126  &   $F_{77}$   &   1.06468   &   1.83767  &   $F_{111}$ &--20.29450 &--32.46123  \\
$F_{221}$ &  --0.69851   &--2.19918  &   $F_{441}$  &--18.73201   &--23.71962  &   $F_{741}$ &   0.56578 &   2.57073  \\
$F_{771}$ &  --0.97378   &--3.04180  &   $F_{222}$  & --0.53376   & --1.11062  &   $F_{662}$ & --0.60404 & --2.24497  \\
$F_{962}$ &    0.61671   &  2.05526  &   $F_{992}$  & --1.01287   & --2.25675  &   $F_{654}$ &--18.17256 & -18.94290  \\
$F_{954}$ &  --0.17620   &--0.38638  &   $F_{984}$  & --0.01658   &   0.04702  &   $F_{987}$ &   0.86987 &   1.30646  \\
$F_{1111}$ &  50.47669   & 89.63168  &   $F_{2211}$ &   1.26945   &   4.44699  &   $F_{4411}$ & 48.94593 &  71.68647  \\
$F_{7411}$ & --0.80741   &--5.46450  &   $F_{7711}$ &   1.53293   &   5.89090  &   $F_{2221}$ &  0.76478 &   2.29325  \\
$F_{6621}$ &   1.16338   &  5.73913  &   $F_{9621}$ & --1.08209   & --4.08511  &   $F_{9921}$ &  1.46146 &   4.33621  \\
$F_{6541}$ &  48.23448   & 60.63713  &   $F_{9541}$ &   0.40465   &   1.24159  &   $F_{9841}$ &  0.15861 &   0.26676  \\
$F_{9871}$ & --0.63502   &--1.85550  &   $F_{2222}$ &   0.76756   &   2.44329  &   $F_{6622}$ &  0.76815 &   3.16961  \\
$F_{6633}$ & --0.46055   &--1.86075  &   $F_{9622}$ & --1.12641   & --3.26978  &   $F_{9633}$ &  0.03132 &   0.14651  \\
$F_{9922}$ &   2.11522   &  5.37994  &   $F_{9933}$ &   0.39929   &   0.77598  &   $F_{9542}$ &  0.43597 &   1.12997  \\
$F_{8762}$ &   0.01923   &  0.04530  &   $F_{4444}$ &  50.01306   &  72.17993  &   $F_{5544}$ & 48.79060 &  59.28204  \\
$F_{7444}$ & --0.85975   &--6.35131  &   $F_{8544}$ &   0.39771   &   4.59345  &   $F_{7744}$ &  1.37066 &   5.70508  \\
$F_{8754}$ &   0.02280   &--0.04550  &   $F_{8844}$ & --0.53371   & --1.07258  &   $F_{7774}$ &--2.21128 & --6.91109  \\
$F_{8874}$ & --0.14830   &--0.36797  &   $F_{7777}$ &   4.58572   &  11.28388  &   $F_{8877}$ &  2.18342 &   3.67611  \\
\end{tabular}
\end{table}

\begin{table}[ht]
\caption{The relationships of the angular momentum basis $\ket{v^l}$
and the Cartesian basis $\ket{n_x n_y}$ for the overtone of the doubly
degenerated modes ($\nu_2$). See Eqs. (\ref{Eab1}) and (\ref{Eab2}) for definition
of $A_1/A_2$ and $E_a/E_b$ symmetry wavefunction.}
\label{e_over}
\begin{tabular}{rrrrrrrrrr}

$\ket{2^{0};A_1}$ =&
$+\frac{\sqrt{2}}{2}\ket{20}$ & $+\frac{\sqrt{2}}{2}\ket{02}$ \\
$\ket{2^{2};E_a}$ = &
$-\frac{\sqrt{2}}{2}\ket{20}$ & $+\frac{\sqrt{2}}{2}\ket{02}$ \\
$\ket{2^{2};E_b}$ = &
+1$\ket{11}$  \\
\hline
$\ket{3^{3};A_1}$ =&
$+\frac{\sqrt{1}}{2}\ket{30}$ &$-\frac{\sqrt{3}}{2}\ket{12}$ \\
$\ket{3^{3};A_2}$  =&
$-\frac{\sqrt{1}}{2}\ket{03}$ &$+\frac{\sqrt{3}}{2}\ket{21}$ \\
$\ket{3^{1};E_a}$ =&
$+\frac{\sqrt{3}}{2}\ket{30}$ &$+\frac{\sqrt{1}}{2}\ket{12}$ \\
$\ket{3^{1};E_b}$ =&
$+\frac{\sqrt{3}}{2}\ket{03}$ &$+\frac{\sqrt{1}}{2}\ket{21}$ \\
\hline
$\ket{4^{0};A_1}$ =&
$+\frac{\sqrt{6}}{4}\ket{40}$ &$+\frac{\sqrt{6}}{4}\ket{04}$ &
$+\frac{\sqrt{4}}{4}\ket{22}$ \\
$\ket{4^{2};E_a}$ =&
$+\frac{\sqrt{2}}{2}\ket{40}$ & $-\frac{\sqrt{2}}{2}\ket{04}$ \\
$\ket{4^{2};E_b}$ =&
$-\frac{\sqrt{2}}{2}\ket{31}$ & $-\frac{\sqrt{2}}{2}\ket{13}$ \\
$\ket{4^{4};E_a}$ =&
$+\frac{\sqrt{2}}{4}\ket{40}$ & $+\frac{\sqrt{2}}{4}\ket{04}$ &
$-\frac{\sqrt{12}}{4}\ket{22}$ \\
$\ket{4^{4};E_b}$ =&
$+\frac{\sqrt{2}}{2}\ket{31}$ & $-\frac{\sqrt{2}}{2}\ket{13}$ \\
\hline
$\ket{5^{3};A_1}$ =&
$+\frac{\sqrt{5}}{4}\ket{50}$ & $-\frac{\sqrt{2}}{4}\ket{32}$ &
$-\frac{\sqrt{9}}{4}\ket{14}$ \\
$\ket{5^{3};A_2}$ =&
$-\frac{\sqrt{5}}{4}\ket{05}$ & $+\frac{\sqrt{2}}{4}\ket{23}$ &
$+\frac{\sqrt{9}}{4}\ket{41}$ \\
$\ket{5^{1};E_a}$ =&
$\frac{\sqrt{10}}{4}\ket{50}$ & $+\frac{\sqrt{4}}{4}\ket{32}$ &
$+\frac{\sqrt{2}}{4}\ket{14}$\\
$\ket{5^{1};E_b}$ =&
$\frac{\sqrt{10}}{4}\ket{05}$ & $+\frac{\sqrt{4}}{4}\ket{23}$ &
$+\frac{\sqrt{2}}{4}\ket{41}$\\
$\ket{5^{5};E_a}$ =&
$\frac{\sqrt{1}}{4}\ket{50}$ & $-\frac{\sqrt{10}}{4}\ket{23}$ &
$+\frac{\sqrt{5}}{4}\ket{14}$\\
$\ket{5^{5};E_b}$ =&
$-\frac{\sqrt{1}}{4}\ket{05}$ & $+\frac{\sqrt{10}}{4}\ket{23}$ &
$-\frac{\sqrt{5}}{4}\ket{41}$\\
\hline
$\ket{6^{0};A_1}$ =&
$+\frac{\sqrt{5}}{4}\ket{60}$ & $+\frac{\sqrt{5}}{4}\ket{06}$ &
$+\frac{\sqrt{3}}{4}\ket{42}$ & $+\frac{\sqrt{3}}{4}\ket{24}$ \\
$\ket{6^{3};A_1}$ =&
$+\frac{\sqrt{2}}{8}\ket{60}$ & $-\frac{\sqrt{2}}{8}\ket{06}$ &
$-\frac{\sqrt{30}}{8}\ket{42}$ & $+\frac{\sqrt{30}}{8}\ket{24}$ \\
$\ket{6^{3};A_2}$ =&
$+\frac{\sqrt{3}}{4}\ket{51}$ & $-\frac{\sqrt{3}}{4}\ket{15}$ &
$+\frac{\sqrt{7}}{4}\ket{33}$ \\
$\ket{6^{2};E_a}$ =&
$+\frac{\sqrt{30}}{8}\ket{60}$ & $-\frac{\sqrt{30}}{8}\ket{06}$ &
$+\frac{\sqrt{2}}{8}\ket{42}$ & $-\frac{\sqrt{2}}{8}\ket{24}$\\
$\ket{6^{2};E_b}$ =&
$-\frac{\sqrt{5}}{4}\ket{51}$ & $-\frac{\sqrt{5}}{4}\ket{15}$ &
$-\frac{\sqrt{6}}{4}\ket{33}$ \\
$\ket{6^{4};E_a}$ =&
$+\frac{\sqrt{3}}{4}\ket{60}$ & $+\frac{\sqrt{3}}{4}\ket{06}$ &
$-\frac{\sqrt{5}}{4}\ket{42}$ & $-\frac{\sqrt{5}}{4}\ket{24}$\\
$\ket{6^{4};E_b}$ =&
$+\frac{\sqrt{2}}{2}\ket{51}$ & $-\frac{\sqrt{2}}{2}\ket{15}$ \\
\hline
$\ket{7^{3};A_1}$ =&
$+\frac{\sqrt{21}}{8}\ket{70}$  & $-\frac{\sqrt{1}}{8}\ket{52}$ &
$-\frac{\sqrt{15}}{8}\ket{34}$   & $-\frac{\sqrt{27}}{8}\ket{16}$ \\
$\ket{7^{3};A_2}$ =&
$-\frac{\sqrt{21}}{8}\ket{07}$  & $+\frac{\sqrt{1}}{8}\ket{25}$ &
$+\frac{\sqrt{15}}{8}\ket{43}$   & $+\frac{\sqrt{27}}{8}\ket{61}$ \\
$\ket{7^{1};E_a}$ =&
$+\frac{\sqrt{35}}{8}\ket{70}$  & $+\frac{\sqrt{15}}{8}\ket{52}$ &
$+\frac{\sqrt{9}}{8}\ket{34}$   & $+\frac{\sqrt{5}}{8}\ket{16}$ \\
$\ket{7^{1};E_b}$ =&
$+\frac{\sqrt{35}}{8}\ket{07}$  & $+\frac{\sqrt{15}}{8}\ket{25}$ &
$+\frac{\sqrt{9}}{8}\ket{43}$   & $+\frac{\sqrt{5}}{8}\ket{61}$ \\
$\ket{7^{5};E_a}$ =&
$+\frac{\sqrt{7}}{8}\ket{70}$  & $-\frac{\sqrt{27}}{8}\ket{52}$ &
$-\frac{\sqrt{5}}{8}\ket{34}$   & $+\frac{\sqrt{25}}{8}\ket{16}$ \\
$\ket{7^{5};E_b}$ =&
$-\frac{\sqrt{7}}{8}\ket{07}$  & $+\frac{\sqrt{27}}{8}\ket{25}$ &
$+\frac{\sqrt{5}}{8}\ket{43}$   & $-\frac{\sqrt{25}}{8}\ket{61}$ \\
$\ket{7^{7};E_a}$ =&
$+\frac{\sqrt{1}}{8}\ket{70}$  & $-\frac{\sqrt{21}}{8}\ket{52}$ &
$+\frac{\sqrt{35}}{8}\ket{34}$   & $-\frac{\sqrt{7}}{8}\ket{16}$ \\
$\ket{7^{7};E_b}$ =&
$-\frac{\sqrt{1}}{8}\ket{07}$  & $+\frac{\sqrt{21}}{8}\ket{25}$ &
$-\frac{\sqrt{35}}{8}\ket{43}$   & $+\frac{\sqrt{7}}{8}\ket{61}$ \\
\hline
$\ket{8^{0};A_1}$ =&
$+\frac{\sqrt{70}}{16}\ket{80}$  & $+\frac{\sqrt{70}}{16}\ket{08}$ &
$+\frac{\sqrt{40}}{16}\ket{62}$   & $+\frac{\sqrt{40}}{16}\ket{26}$ &
$+\frac{\sqrt{36}}{16}\ket{44}$ \\
$\ket{8^{6};A_1}$ =&
$+\frac{\sqrt{1}}{4}\ket{80}$  & $-\frac{\sqrt{1}}{4}\ket{08}$ &
$-\frac{\sqrt{7}}{4}\ket{62}$   & $+\frac{\sqrt{7}}{4}\ket{26}$ \\
$\ket{8^{6};A_2}$ =&
$+\frac{\sqrt{18}}{8}\ket{71}$  & $+\frac{\sqrt{18}}{8}\ket{17}$ &
$-\frac{\sqrt{14}}{8}\ket{53}$   & $-\frac{\sqrt{14}}{8}\ket{35}$ \\
$\ket{8^{2};E_a}$ =&
$+\frac{\sqrt{7}}{4}\ket{80}$  & $-\frac{\sqrt{7}}{4}\ket{08}$ &
$+\frac{\sqrt{1}}{4}\ket{62}$   & $-\frac{\sqrt{1}}{4}\ket{26}$ \\
$\ket{8^{2};E_b}$ =&
$-\frac{\sqrt{14}}{8}\ket{71}$  & $-\frac{\sqrt{14}}{8}\ket{17}$ &
$-\frac{\sqrt{18}}{8}\ket{53}$   & $-\frac{\sqrt{18}}{8}\ket{35}$ \\
$\ket{8^{4};E_a}$ =&
$+\frac{\sqrt{14}}{8}\ket{80}$  & $+\frac{\sqrt{14}}{8}\ket{08}$ &
$-\frac{\sqrt{8}}{8}\ket{62}$   & $-\frac{\sqrt{8}}{8}\ket{26}$ &
$-\frac{\sqrt{20}}{8}\ket{44}$ \\
$\ket{8^{4};E_b}$ =&
$+\frac{\sqrt{7}}{4}\ket{71}$  & $-\frac{\sqrt{7}}{4}\ket{17}$ &
$+\frac{\sqrt{1}}{4}\ket{53}$   & $-\frac{\sqrt{1}}{4}\ket{35}$ \\
$\ket{8^{8};E_a}$ =&
$+\frac{\sqrt{2}}{16}\ket{80}$  & $+\frac{\sqrt{2}}{16}\ket{08}$ &
$-\frac{\sqrt{56}}{16}\ket{62}$   & $-\frac{\sqrt{56}}{16}\ket{26}$ &
$+\frac{\sqrt{140}}{16}\ket{44}$ \\
$\ket{8^{8};E_b}$ =&
$-\frac{\sqrt{1}}{4}\ket{71}$  & $+\frac{\sqrt{1}}{4}\ket{17}$ &
$+\frac{\sqrt{7}}{4}\ket{53}$   & $-\frac{\sqrt{7}}{4}\ket{35}$ \\
\end{tabular}
\end{table}
\begin{table}[h]
\caption{Comparison of {\it ab initio}$^a$ and fitted quadratic force constants for CF$_4$. (Units are consistent with aJ, \AA, and radian.).}
\label{cf4_cst}
\begin{tabular} {lrrrr}
Force Constant & {\it ab initio} &  {\it ab initio} &  Fitted$^c$  & $\sigma^d$\\
               & raw$^a$             &  adjusted$^b$        &       & \\
\hline
$r_e$           &   1.31919 &  1.31526 &   1.31526 &\\ 
$F_{11}      $  &   9.54112 &  9.50679 &   9.50711 & 0.0\% \\
$F_{22}      $  &   1.25733 &  1.24913 &   1.12477 & -10.0\% \\
$F_{33}      $  &   6.32376 &  6.13519 &   6.18043 & +0.7\% \\
$F_{34}      $  &  -1.03160 & -1.03126 &  -1.04559 & -1.4\% \\
$F_{44}      $  &   1.84886 &  1.83767 &   1.82662 & -0.6\% \\
\end{tabular}
\noindent{$^a$  Unadjusted CCSD(T)/cc-pVTZ values}

\noindent{$^b$ Bond distance and diagonal quadratics adjusted to 
reproduce experimental $r_0$ and $\nu_i$}

\noindent{$^c$ bond distance held constants; all quadratic force
constants refined in fit againt complete experimental data set.}

\noindent{$^d$  
Relative deviation of the fitted force constants from
the adjusted {\it ab initio} force constants.}

\end{table}

\begin{table}[ht]
\caption{
 Comparison of experimental and CVPT$^a$
 band origins (cm$^{-1}$) for CF$_4$.
 The fit quadratic force constants of Table \ref{cf4_cst} plus the {\it ab initio}
 cubic and quartic force contants of Table \ref{tab_ff}
 are used for the CVPT calculation.}
\label{tab_cf4}
\squeezetable
\begin{tabular} {cclccrrrll}
$N^a$ & Sym & Obs$^b$ & Uncertainty & Ref. & Obs$-E(2)$ & Obs$-E(4)$ & Obs$-E(6)$
 & \multicolumn{2}{c}{$v_1 v_2 v_3 v_4(c_i^2)^d$}\\
\hline
&&$^{12}$CF$_4$ &&\\
\hline
 2& $E   $&  435.399   &   1  & \onlinecite{Eshe81JMS} &   .227  &  .100 &  .103  & 0100(100\%)& \\
 3& $F_2 $&  631.0593  &   1  & \onlinecite{Bouj98MP}  &   .127  & -.039 & -.031  & 0001(100\%)& \\
 4& $A_1 $&  867.90588 &   1  & \onlinecite{Taby94JRS} &   .476  & -.012 & -.003  & 0200(97\%) & 1000(2\%)\\
 4& $A_1 $&  909.0720  &   1  & \onlinecite{Lolc81JRS} &  -.151  & -.044 & -.045  & 1000(97\%) & 0200(2\%)\\
 5& $F_1 $& 1066.6977  &   1  & \onlinecite{Patt80JMS} &   .300  &  .052 &  .065  & 0101(100\%)& \\
 5& $F_2 $& 1066.1220  &   1  & \onlinecite{Patt80JMS} &   .320  & -.135 & -.112  & 0101(100\%) & \\
 6& $A_1 $& 1261.809   &  10  & \onlinecite{Bouj98MP}  &   .303  & -.184 & -.148  & 0002(90\%) & \\
 6& $E   $& 1262.112   &   1  & \onlinecite{Bouj98MP}  &   .219  & -.244 & -.221  & 0002(100\%)& \\
 6& $F_2 $& 1260.430   &   1  & \onlinecite{Bouj98MP}  &   .186  &  .173 &  .188  & 0002(93\%) & 0010(7\%)\\
 6& $F_2 $& 1283.720   &   1  & \onlinecite{Gaba95JMS} &   .361  & -.005 & -.002  & 0010(93\%) & 0002(7\%) \\
 7& $F_2 $& 1539.3     &  10  & \onlinecite{Jone78JCP} &  -.198  & -.251 & -.241  & 1001(97\%) & 0201(3\%)\\
 8& $F_2 $& 1715.8     &  10  & \onlinecite{Jone78JCP} &  -.225  & -.963 & -.960  & 0110(91\%) & 0102(9\%)\\
 9& $F_2 $& 1889.6$^e$ &  --  & \onlinecite{Jone78JCP} &   .623  &  .887 &  .912  & 0003(84\%) & 0011(16\%)\\
 9& $F_2 $&            &      &                     &         &       &1893.42 & 0003(100\%)& \\
 9& $F_2 $& 1913.2     &  10  & \onlinecite{Jone78JCP} &  2.031  & 1.050 & 1.100  & 0011(84\%) & 0003(16\%)\\
10& $F_2 $& 2168.5     &  10  & \onlinecite{Jone78JCP} &   .674  &  .604 &  .621  & 1002(89\%) & 1010(9\%)\\
10& $F_2 $& 2186.1     &  10  & \onlinecite{Jone78JCP} &  -.602  & -.636 & -.632  & 1010(86\%) & 1002(9\%)\\
11& $F_2 $& 2445.59644 &   1  & \onlinecite{Pine82JMS} &  -.340  &  .025 &  .051  & 2001(94\%) & 1201(5\%)\\
12& $A_1 $& 2553.24(858)&  --  &\onlinecite{Gaba95JMS} &  6.225  & 3.072 & 3.215  & 0020(50\%) & 0012(41\%)\\
12& $E   $& 2570.013   &  --  & \onlinecite{Gaba95JMS} &  1.991  & 1.652 & 1.654  & 0020(93\%) & 0012(7\%)\\
12& $F_2 $& 2561.9124  &  --  & \onlinecite{Gaba95JMS} & -2.127  &-3.061 &-3.067  & 0020(86\%) & 0012(13\%)\\
\hline
&&$^{13}$CF$_4$ &&\\
\hline
 3& $F_2 $&  629.2868 &   1 & \onlinecite{McDo80JMS} &  .270  &  .110 &  .118  & 0001(100\%)& \\
 5& $F_2 $& 1064.39   &  10 & \onlinecite{Jone78JCP} &  .408  & -.033 & -.011  & 0101(100\%)& \\
 6& $F_2 $& 1241.7    &  10 & \onlinecite{Jone78JCP} & -.251  &  .014 &  .009  & 0010(91\%) & 0002(9\%)\\
 6& $F_2 $& 1259.75   &  10 & \onlinecite{Jone78JCP} &  .775  &  .138 &  .160  & 0002(91\%) & 0010(9\%)\\
 7& $F_2 $& 1537.4    &  10 & \onlinecite{Jone78JCP} & -.078  & -.116 & -.107  & 1001(97\%) & 0201(3\%)\\
 8& $F_2 $& 1674.7    &  10 & \onlinecite{Jone78JCP} & -.099  & -.081 & -.087  & 0110(93\%) & 0102(7\%)\\
 9& $F_2 $& 1867.     & 100 & \onlinecite{Jone78JCP} & 1.074  & 1.631 & 1.630  & 0011(89\%) & 0003(11\%)\\
 9& $F_2 $& 1888.$^e$ &  -- & \onlinecite{Jone78JCP} & 1.262  &  .279 &  .353  & 0003(100\%)& \\
 9& $F_2 $&           &     &                     &        &       &1889.37 & 0003(89\%) & 0011(11\%)\\
10& $F_2 $& 2145.     & 100 & \onlinecite{Jone78JCP} & -.698  & -.518 & -.530  & 1010(92\%) & 1002(5\%)\\
10& $F_2 $& 2166.3    &  10 & \onlinecite{Jone78JCP} &  .017  & -.232 & -.200  & 1002(92\%) & 1010(5\%)\\
11& $F_2 $& 2443.3    &  10 & \onlinecite{Jone78JCP} & -.512  &  .011 &  .048  & 2001(94\%) & 1201(6\%)\\
12& $F_2 $& 2477.5    &  10 & \onlinecite{Jone78JCP} &-3.086  &-1.556 &-1.565  & 0020(71\%) & 0012(26\%)\\
\hline
&&$^{14}$CF$_4$ &&\\
\hline
 3& $F_2 $&   627.3490 &   1 & \onlinecite{McDo80JMS} &  .275 &   .121  &  .128  & 0001(100\%)& \\
 5& $F_2 $&  1062.57   &  10 & \onlinecite{Jone78JCP} &  .449 &   .019  &  .041  & 0101(100\%)& \\
 6& $F_2 $&  1208.7    &  10 & \onlinecite{Jone78JCP} & -.107 &  -.055  & -.059  & 0010(99\%) & \\
 6& $F_2 $&  1254.95   &  10 & \onlinecite{Jone78JCP} &  .800 &   .387  &  .407  & 0002(99\%) & \\
 7& $F_2 $&  1535.3    &  10 & \onlinecite{Jone78JCP} & -.204 &  -.230  & -.221  & 1001(97\%) & 0201(3\%)\\
 8& $F_2 $&  1641.6    &  10 & \onlinecite{Jone78JCP} & -.038 &  -.239  & -.246  & 0110(99\%) & \\
 9& $F_2 $&  1833.4    &  10 & \onlinecite{Jone78JCP} & 1.846 &  2.023  & 2.030  & 0011(98\%) & 0003(2\%)\\
 9& $F_2 $&  1881.4$^e$&  -- & \onlinecite{Jone78JCP} &  .717 &  -.212  & -.136  & 0003(99\%) & 0011(1\%)\\
 9& $F_2 $&            &     &                     &       &         &1882.11 & 0003(99\%) & \\
10& $F_2 $&  2073.7$^f$&  10 & \onlinecite{Jone78JCP} & 1.721 &  1.109  & 1.098  & 0210(97\%) & 1010(3\%)\\
10& $F_2 $&            &     &                     &       &         &2075.77 & 0210(99\%) &          \\
10& $F_2 $&  2112.     & 100 & \onlinecite{Jone78JCP} & -.683 &  -.538  & -.547  & 1010(97\%) & 0210(3\%)\\
10& $F_2 $&  2161.9    &  10 & \onlinecite{Jone78JCP} &  .398 &   .218  &  .246  & 1002(96\%) & 0202(3\%)\\
12& $F_2 $&  2412.     & 100 & \onlinecite{Jone78JCP} &-3.208 & -2.645  &-2.638  & 0020(97\%) & 0012(2\%)\\
\hline                                 
RMSD& &     &   &  & 0.0179  & 0.083 & 0.083  &  & \\
\end{tabular}

\noindent{$^a$  $N=4v_1 + 2v_2 + 6v_3 + 3v_4$. This polyad number results from
three independent resonances: $\omega_1 \approx 2\omega_2$,
$3\omega_2 \approx \omega_3$, and $\omega_3 \approx 2\omega_4$.}

\noindent{$^b$ The last figure is significant unless uncertainty
in parenthesis is given otherwise.}

\noindent{$^c$ Uncertainty used in the fit corresponds approximately to the experimental
precison. The states excluded from the fit are 3$\nu_4$ and 2$\nu_2$ + $\nu_3$
whose multiple $F_2$ components are
unresolved, and the $2\nu_3$ bands of $^{12}$CF$_4$, which
differs from experimental results.}

\noindent{$^d$ The largest two components in terms of $v_1 v_2 v_3 v_4$
and their percentage ($c_i^2$) in the basis, based on the sixth order calculation.
Only the components with the percentage larger than 1\% are listed.}

\noindent{$^e$ There are two (0003) $F_2$ states.}

\noindent{$^f$ There are two (0210) $F_2$ states.}

\end{table}

\begin{table}[ht]
\caption{Computed (second order transformed CVPT Hamiltonian) and experimentally derived
vibrational spectroscopic constants of SiF$_4$ and CF$_4$ (cm$^{-1}$).}
\squeezetable
\label{tab_tensor}
\begin{tabular} {rrrrrrrrrr}
Constants
& $^{12}$CF$_4$  & $^{12}$CF$_4$  &$^{12}$CF$_4$  &$^{13}$CF$_4$  & $^{14}$CF$_4$
& $^{28}$SiF$_4$ &  $^{28}$SiF$_4$ &$^{29}$SiF$_4$ & $^{30}$SiF$_4$\\
&  &Heenan I\cite{Heenan}&Heenan II\cite{Heenan}&   &   &   & \Ref{McDo82JCP} &   &  \\
\hline
    $\omega_1$ &  921.596 &   910.52   & 913.25   &  921.596 &  921.596  &  806.101 &	 807.1(12)   &  806.101  &   806.101 \\
    $\omega_2$ &  439.665 &   437.15   & 435.37   &  439.665 &  439.665  &  265.199 &	 267(3)      &  265.199  &   265.199 \\
    $\omega_3$ & 1302.510 &   1292.96  & 1302.26  & 1262.587 & 1227.629  & 1044.044 &	 1044.2(12)  & 1034.820  &  1026.177 \\
    $\omega_4$ &  637.681 &   634.35   & 635.83   &  635.548 &  633.406  &  389.306 &	 389.8(9)    &  387.793  &   386.320 \\
$X_{11}$ &  --1.109 &   -0.279   & -0.275   &  --1.109 &  --1.109  &   --.669 &	 -0.57(5)    &   --.669  &    --.669 \\
$X_{12}$ &--.911$^a$&    0.589   &  0.474   &   --.911 &   --.911  &   --.317 &	 -0.6(11)    &   --.317  &    --.317 \\
$X_{13}$ &  --6.253 &    -4.896   & -4.954   &  --5.908 &  --5.668  &  --3.573 & -3.8(3)      &  --3.503  &   --3.444 \\
$X_{14}$ &   --.675 &     0.608   &  0.712   &   --.928 &  --1.087  & .109$^c$ &   +0.64(7)    & .056  &      .013 \\
$X_{22}$ &--.339$^a$&   -0.284   & -0.254   &  --.339 &   --.339  &   --.168 &	 0.0(5)    &   --.168  &    --.168 \\
$X_{23}$ &  --2.372 &   -0.544   & -0.543   &  --2.315 &  --2.264  &   --.989 &	  -1.5(10)  &   --.979  &    --.968 \\
$X_{24}$ &   --.004 &    -0.283   & -0.273  &     .003 &     .007  &     .506 &	   0.0(5)  &     .503  &	.500 \\
$X_{33}$ &  --4.157 &   -5.642   & -5.536   &  --3.917 &  --3.713  &  --2.845 &	  -3.0058(7)   &  --2.779  &   --2.717 \\
$X_{34}$ &--3.960$^b$&  -1.407   & -1.423   &  --3.756 &  --3.594  &  --1.386 &	  -0.5(4)    &  --1.350  &   --1.318 \\
$X_{44}$ &--.129$^b$&   -0.606   & -0.604   &   --.122 &   --.114  & .165$^c$ &	  -0.22(10)  & .163  &      .161 \\
$G_{22}$ & .542$^a$ &    0.286   &  0.257   &     .542 &     .542  &     .391 &	  [0]	     &     .391  &	.391 \\
$G_{33}$ &    3.833 &    4.317   &  4.241   &    3.618 &    3.432  &    1.799 &	  1.7828(6)  &    1.746  &     1.698 \\
$G_{34}$ &   --.518 &   -1.270   & -1.320   &   --.442 &   --.401  &   --.414 &	$\approx-$0.5&   --.377  &    --.345 \\
$G_{44}$ & .025$^b$ &    0.088   &  0.084   &     .025 &     .025  & .000$^c$ &	 -0.05(15)   & .000  &      .001 \\
$T_{23}$ &   --.039 &   0.045    & 0.044    &   --.037 &   --.035  &   --.101 &	  [0]	     &   --.098  &    --.096 \\
$T_{24}$ &   --.037 &  -0.006    &-0.005    &   --.038 &   --.039  &     .005 &	  [0]	     &     .004  &	.003 \\
      $T_{33}$ &     .221 &   0.683    & 0.670    &	.221 &     .222  &     .196 &	  0.20292(10)&     .195  &	.195 \\
      $T_{34}$ & .080$^b$ &   0.391    & 0.394    &	.065 &     .051  &   --.133 &	$\approx-$0.5&   --.131  &    --.129 \\
      $T_{44}$ & .020$^b$ &   0.091    & 0.090    &	.021 &     .021  &     .014 &	  [0]	     &     .014  &	.013 \\
      $S_{34}$ & .157$^b$ &   ---      & ---	  &	.211 &     .243  &     .019 &	  [0]	     &     .027  &	.033 \\
     $K_{1,44}$&        0 &&&	0      &       0   &2.493$^c$ & 	       & 2.465 &     2.438 \\
     $K_{1,22}$&--3.418$^a$ &&&--3.418 &--3.418    &      0   &		     &       0   &	   0 \\
     $K_{3,44}$&--5.223$^b$ &&&--4.524 &--3.941    &      0   &		     &       0   &	   0 \\
    $K_{1,222}$&      0 &&& 0      &      0        &    0.063 &		     &     0.063 &     0.063 \\
\end{tabular}
{\scriptsize
Spectroscopic constants are identical between curvilinear and rectilinear
formalisms except for those affected by resonances:

\noindent $^a$ 
Due to $\omega_1 \approx 2\omega_2$ resonance, in rectilinear formalism:
$X_{12}=-0.606$; $X_{22}=-0.416$;
$G_{22}=0.619$; and $K_{1,22}=-2.908$ cm$^{-1}$.

\noindent $^b$ 
Due to $\omega_3 \approx 2\omega_4$ resonance, in rectilinear formalism:
$X_{34}=-3.916$; $X_{44}=-0.142$; $G_{34}=-0.550$; $G_{44}=0.018$; $T_{34}=0.067$; $T_{44}=0.024$; $S_{34}=0.161$; 
and $K_{3,44}=5.051$ cm$^{-1}$.

\noindent $^c$ 
Due to resonance $\omega_1 \approx 2\omega_4$, in rectilinear formalism:
$X_{14}=-0.231$; $X_{44}=0.250$; 
$G_{44}=-0.085$; and $K_{1,44}=2.699$ cm$^{-1}$.

}
\end{table}

 \begin{table}[ht]
 \caption{
 Comparison of experimental, CVPT$^a$
 and $U$(2) algebraic model band origins (cm$^{-1}$) for SiF$_4$.
 The {\it ab initio} force contants of Table \ref{tab_ff} are used for the CVPT calculation.}
 \label{tab_vib}
 \squeezetable
 \begin{tabular} {cllrrrrll}
$N^a$ & Sym & Obs$^b$ & Obs$-E(2)$ & Obs$-E(4)$ & Obs$-E(6)$  & Obs$-U(2)$$^c$
 & $\nu_1 \nu_2 \nu_3 \nu_4 (c_i^2)^d$\\
\hline
 2& $E   $&  264.2(10)     &  .00   & -.03    & -.03   & -.21  & 0100(100\%)& \\
 3& $F_2 $&  388.4448(2)   & -.01   & -.03    & -.03   & -.41  & 0001(100\%)& \\
 6& $A_1 $&  800.6(3)      & -.24   & -.30    & -.30   & 1.03  & 1000(93\%) & 0002(6\%) \\
 6& $F_2 $&  776.3(5)      & -.81   & -.84    & -.83   &  .97  & 0002(100\%)& \\
 8& $F_2 $& 1031.3968(3)   & -.64   & -.76    & -.75   & 1.72  & 0010(100\%)& \\
 8& $E   $& 1064.2(4)      &  .00   & -.06    & -.06   &  .01  & 1100(92\%)& 0102(7\%)\\
 9& $F_2 $& 1164.2(2)      &  .48   &  .48    &  .51   &  .03  & 0003(90\%)& 1001(10\%)\\
 9& $F_2 $& 1189.7(3)      & -.70   & -.83    & -.82   & 1.07  & 1001(90\%)& 0003(10\%)\\
10& $F_2 $& 1294.05(10)    &  .25   &  .11    &  .11   &  .15  & 0110(100\%)& \\
11& $F_2 $& 1418.75(10)    &  .16   &  .08    &  .08   &  .22  & 0011(100\%)& \\
14& $F_2 $& 1804.5(1)$^e$  & 1.05   &  .70    &  .72   & -.21  & 0012(97\%)& 1010(2\%)\\
14& $F_2 $&                &1805.34 &1805.49  &1805.48 &       & 0012(96\%)& 1010(4\%)\\
14& $F_2 $&                &1807.20 &1807.38  &1807.37 &       & 0012(100\%)& \\
14& $F_2 $& 1828.17(2)     & -.56   & -.65    & -.65   & -.57  & 1010(93\%)& 0012(7\%)\\
16& $F_2 $& 2059.1(3)      & -.04   & -.17    & -.18   & 1.09  & 0020(100\%)& \\
20& $F_2 $& 2602.55(10)$^f$& -.14   & -.71    & -.67   &-1.24  & 1012(70\%)& 2010(12\%)\\
20& $F_2 $&                &2604.97 &2605.28  &2605.09 &       & 1012(81\%)& 0014(9\%)\\
20& $F_2 $&                &2606.96 &2607.20  &2607.04 &       & 1012(86\%)& 0014(14\%)\\
20& $F_2 $& 2623.8(1)      &-1.30   &-1.47    &-1.43   &  .12  & 2010(83\%)& 1012(15\%)\\
24& $F_2 $& 3068.5(1)      & -.74   &-1.71    &-1.68   & -.63  & 0030(100\%)& \\
\hline
RMSD&     &                & 0.59   & 0.74    & 0.73  & 0.79   \\
\end{tabular}

\noindent{$^a$  $N=6v_1 + 2v_2 + 8v_3 + 3v_4$. This polyad number results from
three independent resonances: $\omega_1+\omega_2 \approx \omega_3$,
$\omega_1 \approx 2\omega_4$, and $\omega_1 \approx 3\omega_2$.}

\noindent{$^b$  The observations are from \Ref{McDo82JCP}.
Standard deviations given in parentheses.}

\noindent{$^c$  The $U$(2) algebraic model calculations are from \Ref{Hou98CPL}.}

\noindent{$^d$ The largest two components in terms of $v_1 v_2 v_3 v_4$
and their percentage ($c_i^2$) in the basis, based on the sixth order calculation.
Only the components with the percentage larger than 1\% are listed.}

\noindent{$^e$ There are three (0012) $F_2$ states.}

\noindent{$^f$ There are three (1012) $F_2$ states.}

\end{table}


\begin{thebibliography}{99}

\bibitem{KirkOthmer} {\it Kirk-Othmer encyclopedia of chemical
technology}, 4th Ed., R. E. Kirk, D. F. Othmer, J. I. Kroschwitz, Eds.
(Wiley, New York, 1998)
\bibitem{blood}  H.-G. Mack and H. Oberhammer, J. Chem. Phys. {\bf 87}, 2158 (1987)
and references therein


\bibitem{Kim96} J.-H. Kim, S.-H. Seo, S.-M. Yun, H.-Y. Chang,
K.-M. Lee, and C.-K. Choi, J. Electrochem. Soc. {\bf 143}, 2990 (1996)
and references therein.



\bibitem{JAP} J. Ding and N. Hershkowitz, {\rm Appl. Phys. Lett.}
{\bf 68}, 1619 (1996)
%

\bibitem{philips} P. G. M. Sebel, L. J. F. Hermans, and H. C. W. Beijerinck,
{\rm J. Vac. Sci. Tech. A} {\bf 17}, 755 (1999); see also
T. Lagarde, J. Pelletier, and Y. Arnal, {\rm J. Vac. Sci. Tech. A} {\bf 17},
118 (1999)
%


\bibitem{Francis} P. Francis, C. Chaffin, A. Maciejewski, and C. Oppenheimer,
{\rm Geophys. Res. Lett.} {\bf 23}, 249 (1996)

\bibitem{30Si} K. Tanaka, S. Isomura, H. Kaetsu, Y. Yatsurugi, M. Hashimoto,
K. Togashi, and S. Arai, {\rm Bull. Chem. Soc. Japan} {\bf 69}, 493 (1996)

\bibitem{early} B. Monostori and A. Weber, J. Chem. Phys. {\bf 33}, 1867 (1960);
A. Maki, E. K. Plyler, and R. Thibault, J. Chem. Phys. {\bf 37}, 1899 (1960);
P. J. H. Woltz and A. H. Nielsen, J. Chem. Phys. {\bf 20}, 307 (1952);

\bibitem{Jon78b} L. H. Jones, B. J. Krohn, and R. C. Kennedy, J. Mol. Spectrosc.
{\bf 70}, 288 (1978)

\bibitem{Jone78JCP}
L.~H. Jones, C.~Kennedy, and S.~Ekberg,
\newblock J. Chem. Phys.  {\bf 69},  833   (1978).



\bibitem{Jea73} A. C. Jeannotte II, D. Legler, and J. Overend,
Spectrochim. Acta A {\bf 29}, 1915 (1973)


\bibitem{Wittig} J. J. Tiee and C. Wittig, Appl. Phys. Lett. {\bf 30}, 420 (1977);
J. Appl. Phys. {\bf 49}, 61 (1978)

\bibitem{Eshe81JMS}
P.~Esherick, A.~Owyoung, and C.~W. Patterson,
\newblock J. Molec. Spectrosc.  {\bf 86},  250   (1981).

\bibitem{Lolc81JRS}
{J. E. Lolck},
\newblock J. Raman Spectrosc.  {\bf 11},  294   (1981).

\bibitem{Taby94JRS}
A.~Tabyaoui, B.~Lavorel, R.~Saint-Loup, and M.~R\"otger,
\newblock J. Raman Spectrosc.  {\bf 25},  255   (1994).

\bibitem{Patt80JMS}
C.~W. Patterson, R.~S. McDowell, N.~G. Nereson, R.~F. Begley, H.~W. Galbraith,
  and B.~J. Krohn,
\newblock J. Molec. Spectrosc.  {\bf 80},  71   (1980).


\bibitem{Pouv2+v4} G. Poussigue, G. Tarrago, and A. Valentin, J. Phys. B.
{\bf 47}, 1155 (1986)

\bibitem{tar81} G. Tarrago, G. Poussigue, M. Dang-Nhu, and J. Kauppinen,
J. Mol. Spectrosc. {\bf 86}, 232 (1981)

\bibitem{McDo80JMS}
R.~S. McDowell, M.~J. Reisfeld, H.~W. Galbraith, B.~J. Krohn, H.~Flicker,
R.~C. Kennedy, J.~P. Aldridge, and N.~G. Nereson,
\newblock J. Molec. Spectrosc.  {\bf 83},  440   (1980).

\bibitem{Pine82JMS}
A.~S. Pine, \newblock J. Molec. Spectrosc.  {\bf 96},  395   (1982)

\bibitem{Dang81JMS}
M. Dang-Nhu, G. Graner, and G. Guelachvili, J. Mol. Spectrosc. {\bf 85}, 358 (1981)



\bibitem{tak81} M. Takami, J. Chem. Phys. {\bf 73}, 2665 (1980)

\bibitem{Gaba95JMS}
T.~Gabard, A.~Nikitin, J.~P. Champion, G.~Pierre, and A.~S. Pine,
\newblock J. Molec. Spectrosc.  {\bf 170},  431   (1995).

\bibitem{e}  T. Gabard, L. Pierre, and M. Takami, Mol. Phys. {\bf 85}, 735 (1995)

\bibitem{Duncan} J. L. Duncan and I. M.  Mills, Spectrochim. Acta {\bf 20}, 1089 (1964)

\bibitem{Chalmers} A. A. Chalmers and D. C. McKean, Spectrochim. Acta {\bf 22}, 251 (1966)

\bibitem{Jea78} A. C. Jeannotte II, C. Marcott, and J. Overend, J. Chem. Phys.
{\bf 65}, 2076 (1978)

\bibitem{Bro91} S. Brodersen, J. Mol. Spectrosc. {\bf 145}, 331 (1991)

\bibitem{Lars93JMS}
S.~G. Larsen and S.~Brodersen,
\newblock J. Molec. Spectrosc.  {\bf 157},  220   (1993).

\bibitem{Bouj98MP}
V.~Boujut, F.~Michelot, and C.~Leroy,
\newblock Mol. Phys.  {\bf 93},  879   (1998).


\bibitem{stds}
V. G. Tyuterev, Y. L. Babikov, S. A. Tashkun, V. I. Perevalov, A. Nikitin,
J. P. Champion, C. Wenger, G. Pierre, J. C. Hilico, and M. Lo\"ete,
J. Quant. Spec. Radiat. Transfer {\bf 52}, 459 (1994) based on \Ref{e}

\bibitem{Mari95JCP}
W.~Maring, J.~P. Toennies, R.~G. Wang, and H.~B. Levene,
\newblock J. Chem. Phys.  {\bf 103},  1333   (1995).

\bibitem{Heenan} R. K. Heenan, Ph.D. Thesis (U. of Reading, UK, 1979)



\bibitem{Hech60JMS}
K.~T. Hecht,
\newblock J. Molec. Spectrosc.  {\bf 5},  355   (1960).



\bibitem{McDo82JCP}
R.~S. McDowell, M.~J. Reisfeld, C.~W. Patterson, B.~J. Krohn, M.~C. Vasquez,
  and G.~A. Laguna,
\newblock J. Chem. Phys.  {\bf 77},  4337   (1982).


\bibitem{Hou98CPL}
X.-W. Hou, S.-H. Dong, M.~Xie, and Z.-Q. Ma,
\newblock Chem. Phys. Lett.  {\bf 283},  174   (1998).

\bibitem{Hou98AP} 
X.-W. Hou, M. Xie, S.-H. Dong, and Z.-Q. Ma, 
\newblock Ann. Phys. (NY) {\bf 203}, 340 (1998).


\bibitem{IachelloLevine} F. Iachello, Chem. Phys. Lett. {\bf 78}, 581 (1981);
F. Iachello and R. D. Levine, J. Chem. Phys. {\bf 77}, 3046 (1982)


\bibitem{Pat82}
C. W. Patterson and A. S. Pine, J. Mol. Spectrosc.
{\bf 96}, 404 (1982)


\bibitem{Taka83JCP} M. Takami and H. Kuze, J. Chem. Phys. {\bf 78}, 2204 (1983).

\bibitem{Jori89CJP} L. J\"orissen, H. Prinz, W. A. Kreiner,
C. Wenger, G. Pierre, G. Magerl, W. Schupita, 
Can. J. Phys. {\bf 67}, 532 (1989).

\bibitem{Demaison}
J. Breidung, J. Demaison, L. Margules, and W. Thiel, Chem. Phys. Lett., 
submitted (``Equilibrium Structure of SiF4'');
preprint communicated to authors



\bibitem{sif4} J. M. L. Martin and P. R. Taylor, {\rm J. Phys. Chem.
A} {\bf 103}, 4427 (1999).

\bibitem{Joh86} G. K. Johnson, {\rm J. Chem. Thermodyn.} {\bf 18}, 801 (1986)




\bibitem{ch4}
T. J. Lee, J. M. L. Martin, and P. R. Taylor, {\rm J. Chem. Phys.} {\bf 102},
254 (1995)

\bibitem{carter} %
S. Carter, H. M. Shnider, and J. M. Bowman, {\rm J. Chem. Phys.}
{\bf 110}, 8417 (1999)

\bibitem{Venu99JCP}
E.~Venuti, L.~Halonen, and R.~G.~Della Valle,
\newblock J. Chem. Phys.  {\bf 110},  7339   (1999).

\bibitem{Wang99JCP}
X.-G. Wang and E.~L. Sibert,
\newblock J. Chem. Phys.  {\bf 111},  4510   (1999).




\bibitem{molpro} H.-J. Werner and P. J. Knowles,
MOLPRO 97.3, a package of {\em ab initio} programs,
with contributions from
J.  Alml\"of, R. D.  Amos, A.  Berning, D. L.  Cooper,
M. J. O.  Deegan, A. J.  Dobbyn,
F.  Eckert, S. T.  Elbert, C.  Hampel, R.  Lindh,
A. W.  Lloyd, W.  Meyer, A.  Nicklass,
K. A.  Peterson, R. M.  Pitzer, A. J.  Stone,
P. R.  Taylor, M. E.  Mura, P.  Pulay,
M.  Sch\"utz, H.  Stoll  and T. Thorsteinsson.


\bibitem{Pur82} G. D. Purvis III and R. J. Bartlett,
{\rm J. Chem. Phys.} {\bf 76}, {1910} (1982)

\bibitem{Rag89}
K. Raghavachari, G. W. Trucks, J. A. Pople, and
M. Head-Gordon, {\rm Chem. Phys. Lett.} {\bf 157}, 479 (1989)

\bibitem{Ham92}
P. J. Knowles, C. Hampel, and H. J. Werner, {\rm J. Chem. Phys.} {\bf 99},
{5219} (1993)



\bibitem{Lee95} T. J. Lee and G. E. Scuseria, in {\it Quantum
mechanical electronic structure calculations with chemical accuracy},
Ed. S. R. Langhoff (Kluwer Academic Publishers, Dordrecht, The
Netherlands, 1995).

\bibitem{Lee89IJQC}
T. J. Lee and P. R. Taylor, Int. J. Quantum Chem. Symp. {\bf 23}, 199 (1989)


\bibitem{Papo82}
D.~Papou{\v{s}}ek and M.~R. Aliev, {\em Molecular Vibrational-Rotational
  Spectra}
\newblock (Elsevier Scientific Publishing Company,: New York  1982).


\bibitem{spectro}
A. Willetts, J. F. Gaw, W. H. Green Jr., and N. C. Handy, {\it
SPECTRO 1.0, a second-order rovibrational perturbation theory program}
(University Chemical Laboratory, Cambridge, UK, 1989)

\bibitem{Gaw90}
J. F. Gaw, A. Willetts, W. H. Green, and N. C. Handy, in {\it Advances
in molecular vibrations and collision dynamics} (ed. J. M. Bowman),
JAI Press, Greenwich, CT, 1990.



\bibitem{Hodg83MP}
D.~P. Hodgkinson, R.~K. Heenan, A.~R. Hoy, and A.~G. Robiette,
\newblock Mol. Phys.  {\bf 48},  193   (1983).

\bibitem{depert}
This is fairly trivially done by taking apart the compound fractions
in Eq.(3.20) of \Ref{Hodg83MP}, then inserting a test that skips the
appropriate one among the four affected terms if the respective
denominator corresponds to the resonant interaction. Alternatively, the
reciprocal denominators may be precomputed and stored in a 3-dimensional
array, and any that correspond to resonant interactions set to zero before
the actual spectroscopy routines are entered, similar to
J. M. L. Martin and P. R. Taylor,
Spectrochim. Acta A {\bf 53}, 1039 (1997).

\bibitem{Dun89} T. H. Dunning, Jr., {\rm J. Chem. Phys.} {\bf 90}, {1007}
(1989)

\bibitem{Woo93} D. E. Woon and T. H. Dunning Jr.,
{\rm J. Chem. Phys.} {\bf 98}, {1358} (1993).

\bibitem{sio} J. M. L. Martin and O. Uzan,
{\rm Chem. Phys. Lett.} {\bf 282}, {16} (1998)


\bibitem{so2}
J. M. L. Martin, {\it J. Chem. Phys.} {\bf 108}, {2791} (1998)

\bibitem{so3}
J. M. L. Martin, {\it Spectrochim. Acta A} {\bf 55}, {709} (1999)
(special issue ``Theoretical spectroscopy: state of the science'')

\bibitem{Ken92} R. A. Kendall, T. H. Dunning Jr., and R. J. Harrison,
{\it J. Chem. Phys.} {\bf 96}, 6796 (1992).

\bibitem{Mar94CPL}
J. M. L. Martin and P. R. Taylor,
Chem. Phys. Lett. {\bf 225}, 473 (1994).

\bibitem{Mar94JPC}
J. M. L. Martin, J. P. Fran\c{c}ois, and R. Gijbels,
J. Phys. Chem. {\bf 98}, 11394 (1994).

\bibitem{Mar98CPL}
J. M. L. Martin,
Chem. Phys. Lett. {\bf 292}, 411 (1998)

\bibitem{sih4} J. M. L. Martin, K. K. Baldridge, and T. J. Lee,
Mol. Phys. {\bf xx}, yyy (1999)



\bibitem{Pick72JCP}
H.~M. Pickett,
\newblock J. Chem. Phys.  {\bf 56},  1715   (1972).

\bibitem{Sibe88JCP}
{E. L. Sibert},
\newblock J. Chem. Phys.  {\bf 88},  4378   (1988).

\bibitem{Sibe88CPC}
{E. L. Sibert},
\newblock Comp. Phys. Comm.  {\bf 51},  149   (1988).

\bibitem{Vall99JCC}
R.~G.~Della Valle, L.~Halonen, and E.~Venuti,
\newblock J. Comput. Chem.,
\newblock (in press).

\bibitem{Halo88CPC}
{L. Halonen and M. S. Child},
\newblock Comp. Phys. Comm.  {\bf 51},  173   (1988).

\bibitem{Halo97JCP}
L.~Halonen,
\newblock J. Chem. Phys.  {\bf 106},  831   (1997).

\bibitem{Grif61}
J.~S. Griffith,
\newblock {\em The Theory of Transition-Metal Ions},
\newblock Cambridge University Press, Cambridge, (1961),
\newblock Table A20 of Appendix 2.

\bibitem{Tana54JPSJ}
Y.~Tanabe and S.~Sugano,
\newblock J. Phys. Soc. Jap.  {\bf 9},  753   (1954).

\bibitem{Paul35}
L.~Pauling and E.~B.Wilson, {\em Introduction to Quantum Mechanics with
  Applications to Chemistry}
\newblock (Dover Publications, Inc.: New York  1935).

\bibitem{Shaf44RMP}
W.~H. Shaffer,
\newblock Rev. Mod. Phys.  {\bf 16},  245   (1944).

\bibitem{Cohe77}
C.~Cohen-Tannoudji, B.~Diu, and F.~Lalo{\"{e}},
\newblock {\em Quantum Mechanics}, vol~1, pp 727-741,
\newblock Wiley, New York, (1977).

\bibitem{Pak97JCP} Y.~Pak, E.~L. Sibert~III, and R.~C. Woods, 
\newblock J. Chem. Phys. {\bf 107}, 1717 (1997).

\bibitem{Maple}
{MAPLE V Release 4, Waterloo Maple Software, Waterloo, Ontario, 1981-1994}.

\bibitem{Lemu94JCP}
R.~Lemus and A.~Frank,
\newblock J. Chem. Phys.  {\bf 101},  8321   (1994).

\bibitem{Ma96PRA}
Z.-Q. Ma, X.-W. Hou, and M.~Xie,
\newblock Phys. Rev. A  {\bf 53},  2173   (1996).

\bibitem{Fran99JMS}
A. Frank, R. Lemus, F. P\'{e}rez-Bernal, and R. Bijker, 
\newblock J. Mol. Spectrosc. {\bf 196}, 329 (1999).

\bibitem{Sibe96JCP}
A. B. McCoy and E. L. Sibert III,
\newblock J. Chem. Phys. {\bf 105}, 459 (1996);
E. L. Sibert III and A. B. McCoy,
\newblock J. Chem. Phys. {\bf 105}, 469 (1996).

\bibitem{Jahn38PRS1}
H.~A. Jahn,
\newblock Proc. Roy. Soc.  {\bf A168},  469   (1938).

\bibitem{Jahn35APL}
H.~A. Jahn,
\newblock Ann. Phys. Lpz.  {\bf 23},  529   (1935).

\bibitem{Jahn38PRS2}
H.~A. Jahn,
\newblock Proc. Roy. Soc.  {\bf A168},  495   (1938).

\bibitem{Gray79MP}
D.~L. Gray and A.~G. Robiette,
\newblock Mol. Phys.  {\bf 37},  1901   (1979).

\bibitem{Mill58MP}
I.~M. Mills,
\newblock Mol. Phys.  {\bf 1},  107   (1958),
\newblock Fig. 1 of this work is inconsistent with its definition of symmetry
  coordinates.

\bibitem{Robi76MP}
A.~G. Robiette, D.~L. Gray, and F.~W. Birss,
\newblock Mol. Phys.  {\bf 32},  1591   (1976).


\bibitem{McCo92MP}
{A. B. McCoy and E. L. Sibert},
\newblock Mol. Phys.  {\bf 77},  697   (1992).

\bibitem{McCo91JCP1}
A.~B. McCoy and E.~L. Sibert,
\newblock J. Chem. Phys.  {\bf 95},  3476   (1991).

\bibitem{McCo91JCP2}
{A. B. McCoy, D. C. Burleigh, and E. L. Sibert},
\newblock J. Chem. Phys.  {\bf 95},  7449   (1991).

\bibitem{Mard98JMS}
{K. L. Mardis and E. L. Sibert},
\newblock J. Molec. Spectrosc.  {\bf 187},  167   (1998).


\end{thebibliography}
\end{document}